\documentclass[aps,prd,eqsecnum,twocolumn]{revtex4}

\usepackage{graphicx}
\usepackage{lscape}
\usepackage{indentfirst}
\usepackage{latexsym}
\usepackage{multirow}
\usepackage{tabls}

\usepackage{amsfonts}
\usepackage{amsmath}
\usepackage{bm}
\usepackage{mathrsfs}

\def\arreq{ \!\! = \!\! }

\def\a{ \alpha }
\def\b{ \beta }

\def\ve{ \varepsilon }

\def\ve{ \varepsilon }

\def\barh{ {\bar h} }

\def\cB{ {\cal B} }
\def\cC{ {\cal C} }
\def\cE{ {\cal E} }

\def\cR{ {\cal R} }

\def\bk{ {\bf k} }

\def\dx{ { \dot{x} } }
\def\dz{ { \dot{z} } }

\def\hata{ {\hat{a}} }
\def\hatb{ {\hat{b}} }
\def\hatc{ {\hat{c}} }
\def\hatd{ {\hat{d}} }

\def\hath{ {\hat{h}} }
\def\hati{ {\hat{i}} }
\def\hatj{ {\hat{j}} }

\def\hatm{ {\hat{m}} }
\def\hatn{ {\hat{n}} }

\def\hats{ {\hat{s}} }
\def\hatt{ {\hat{t}} }

\def\hatzero{ {\hat{0}} }

\def\b0{ {\bf 0} }

\def\lsim{\mathrel{\rlap{\lower3pt\hbox{\hskip1pt$\sim$}}
    \raise1pt\hbox{$<$}}}                
\def\gsim{\mathrel{\rlap{\lower3pt\hbox{\hskip1pt$\sim$}}
    \raise1pt\hbox{$>$}}}         

\def\coordeq{ \, \mathrel{ \rlap{\hbox{\hskip-2.5pt$=$} }
    \raise4pt\hbox{$\cdot$}} \, }                

\begin{document}

\title{Self-force on extreme mass ratio inspirals
via curved spacetime effective field theory}

\author{Chad R. Galley$^1$\footnote{crgalley@umd.edu} and B. L. Hu$^{1,2}$\footnote{blhu@umd.edu}}


\affiliation{$^1$Maryland Center for Fundamental Physics, Department
of Physics, University of Maryland, College Park, Maryland, 20742 }

\affiliation{$^2$Perimeter Institute for Theoretical Physics, 31
Caroline Street North, Waterloo, Ontario N2L 2Y5 Canada}


\begin{abstract}
In this series we construct an effective field theory (EFT) in
curved spacetime to study gravitational radiation and
backreaction effects. We begin in this paper with a derivation of the
self-force on a compact object moving in the background spacetime of
a supermassive black hole. The EFT approach utilizes the disparity
between two length scales, which in this problem are the size of
the compact object $r_m$ and the radius of curvature of the
background spacetime ${\cal R}$ such that $\varepsilon \equiv r_m /
{\cal R} \ll 1$, to treat the orbital dynamics of the compact object,
described as an effective point particle, separately from its tidal
deformations. The equation of motion of an effective relativistic
point particle coupled to the gravitational waves generated by its
motion in a curved background spacetime can be derived without making
a slow motion or weak field approximation, as was assumed in earlier
EFT treatment of post-Newtonian binaries. Ultraviolet divergences are
regularized using Hadamard's {\it partie finie} to isolate the
non-local finite part from the quasi-local divergent part. The latter
is constructed from a momentum space representation for the graviton
retarded propagator and is evaluated using dimensional regularization
in which only logarithmic divergences are relevant for renormalizing
the parameters of the theory. As a first important application of
this framework we explicitly derive the first order self-force given
by Mino, Sasaki, Tanaka, Quinn and Wald. Going beyond the point
particle approximation, to account for the finite size of the object,
we demonstrate that for extreme mass ratio inspirals the motion of a
compact object is affected by tidally induced moments at
$O(\varepsilon^4)$, in the form of an Effacement Principle. The
relatively large radius-to-mass ratio of a white dwarf star allows
for these effects to be enhanced until the white dwarf becomes
tidally disrupted, a potentially $O(\varepsilon^2)$ process, or
plunges into the supermassive black hole.
This work provides a new foundation for further exploration of higher
order self force corrections, gravitational radiation and spinning compact objects.

\end{abstract}

\maketitle



\section{ Introduction and  Main Points}

In two previous papers
\cite{GalleyHu:PRD72,GalleyHuLin:PRD74,Galley:PhD}, using a
stochastic field theory approach based on open system concepts,  we
derive the scalar, electromagnetic and gravitational self-force to
leading order on a particle moving in an arbitrary curved background
spacetime. 
We begin with the particle following a quantum mechanical path while
interacting with a linear quantum field
\cite{JohnsonHu:PRD65,JohnsonHu:FoundPhys35,Johnson:PhD}. The conditions on a stochastic
field theory (for open systems) to emerge from a quantum field theory
(of closed systems) are that the mass and size of the particle are
large enough so the particle worldline is sufficiently decohered from
its interactions with the quantum fluctuations of the field that it
can be considered as quasi-classical, and yet  sufficiently small
that quantum fluctuations manifest as classical stochastic forces
\cite{GellMannHartle:PRD47}.

\subsection{Quantum, Stochastic and Effective Field Theories}

When there is a significant discrepancy between the two mass (or
energy or length) scales in a problem, as in the extreme mass ratio
binary systems under consideration here, one could use an open system
stochastic description for their dynamics, such as developed in
\cite{GalleyHu:PRD72,GalleyHuLin:PRD74}. When this discrepancy is
huge (such as between the QCD and GUT scales in particle physics) the
stochastic component is strongly suppressed in the wide range between
the two scales, away from the threshold region
\cite{CalzettaHu:PRD55}. Then the
stochastic field theory description will give rise to an effective
field theory (EFT) description \cite{Burgess:EFT} for the motion of
the small mass subsystem. Due to the large separation in the mass
scales quantum loop corrections from the field and the intrinsic
quantum mechanical worldline fluctuations are very strongly
suppressed. These two factors render a quantum field theory (QFT)
into a stochastic field theory  (with sufficiently decohered
histories) and in turn (with sufficiently small stochasticity) an
effective field theory for the dynamics of the reduced systems. We
shall explain the essence and demonstrate the advantages in taking a
field theory approach to treat radiation-reaction of classical
systems.
The application of EFT to the treatment of gravitational radiation from
post-Newtonian (PN) binary systems was first introduced by Goldberger
and Rothstein \cite{GoldbergerRothstein:PRD73}. Our formulation of a
curved spacetime effective field theory (CS-EFT) goes beyond with two
important features: it is for any curved spacetime background and
there is no slow motion or weak field restrictions \footnote{In
addition to \cite{GalleyHu:PRD72, GalleyHuLin:PRD74,
GoldbergerRothstein:PRD73} for applying effective field theory
techniques to the motion of extended bodies and charges, see also a
recent paper of \cite{KolSmolkin:grqc}.}.

In this and three subsequent papers \cite{Galley:EFT2, Galley:EFT3,
Galley:spin} we construct a CS-EFT and apply it to derive the
self-force on a compact object moving in an arbitrary curved
background  For concreteness, the background spacetime is assumed to
be that of a supermassive black hole (SMBH) with curvature scale
$\cR$ much larger than the size of the compact object $r_m$.
The smallness of  the  ratio $\ve \equiv r_m / \cR$ makes it a good
expansion parameter for  a perturbation theory treatment describing
the extreme mass ratio inspiral (EMRI) of the compact object. These
binary systems are expected to be good candidates for detecting
gravitational wave signatures using the space-based gravitational
wave interferometer LISA \cite{LISA}. We now give some realistic
numbers for astrophysical processes in this category to delimit the
range of validity for carrying out these perturbation expansions  for
these EMRI sources detectable in LISA's bandwidth . However, we
emphasize that the formalism we construct here is of a sufficiently
general nature that it can be applied to any compact object  moving
in an arbitrary curved background, including those spacetimes sourced
by some form of stress-energy and those possessing a cosmological
constant.

\subsection{Relevant scales in EMRIs}

Consider the motion of a compact object (a black hole, neutron star
or white dwarf  with a mass $m$ ranging from  about 1 to 100  solar
masses) moving through the spacetime of a SMBH with a mass $M \sim
10^{5-7} M_\odot$. We have in mind that the compact object moves in a
stationary background provided by the supermassive black hole, such
as the Schwarzschild or Kerr spacetimes. Such spacetimes are
appropriate for a description of the EMRI in which the compact object
is bound by the gravitational pull of the SMBH. By emitting
gravitational waves the binary system loses energy until the compact
object plunges into the SMBH.
The emission of gravitational radiation from such a system is
expected to be detected with the anticipated construction and launch
of the LISA space-based interferometer \cite{LISA}.

It is believed that most SMBHs lurking in the middle of galaxies,
which are thought to host the prime sources of gravitational wave
emissions detectable by LISA, are spinning and clean in the sense
that most, if not all, of the surrounding material has already fallen
into the black hole.  (Active galactic nuclei are a notable exception
\cite{LISAsources}.)  Because of this the Kerr background is perhaps
the most astrophysically relevant spacetime for the extreme mass
ratio inspiral. The Kerr solution is vacuous ($R_{\mu\nu}=0$),
stationary and stable under small perturbations
\cite{Whiting:JMathPhys30} and possesses two Killing fields. The
first Killing field ($\xi^\alpha$) is time-like and describes time-translation
invariance everywhere outside of the ergoregion. The second ($\psi^\alpha$) is
space-like and describes the axial  symmetry  of the
spacetime.

The Ernst \cite{Ernst:JMathPhys17} and Preston-Poisson
\cite{PrestonPoisson:PRD74}  geometries describe  a black hole
immersed in an external magnetic field. From an astrophysical
viewpoint, the external magnetic fields that a black hole at the
center of a galaxy experiences are relatively weak and unlikely to
significantly affect the motion of the compact object until a very
high order in the perturbation theory.

There are two relevant length scales in EMRIs. The smaller scale is
set by the size of the compact object itself, denoted $r_m$. For an
astrophysical black hole its radius is $r_{bh} = 2 G_N m \sim m/
m_{pl}^2$ where $m_{pl}^{-2} = 32\pi G_N$ in units where $\hbar = c=
1$ \footnote{We follow the conventions of \cite{MTW} so that the
metric has mostly positive signature $(-,+,+,+)$.}.  For a neutron star with a mass
$\approx 1.4 M_\odot$ and a radius of $10-16$ km it follows that
$r_{ns} \approx 4.8 -7.7 \, G_N m \sim m/m_{pl}^2$. Therefore, it is
to be expected that the size of the compact object, be it a black
hole or a neutron star, is of the order of its mass  \footnote{See Section \ref{sec:finsize} for more details concerning a white dwarf, which has a typical radius $r_{wd} \sim 10^3 m / m_{pl}^2$.} .

The second relevant scale is the radius of curvature of the
background spacetime, $\cR$. We take $\cR$ to be the
following curvature invariant
\begin{eqnarray}
    \cR = \big( R_{\mu\alpha \nu \beta} R^{\mu\alpha \nu \beta} \big) ^{-1/4}   .  \label{curvinvariant0}
\end{eqnarray}
For a (possibly rotating) stationary SMBH the radius of curvature is
\begin{eqnarray}
    \cR \sim  \sqrt{ \frac{ m_{pl}^2 r^3}{M} }
\end{eqnarray}
where $r$ is the typical orbital distance for the compact object away
from the central black hole. For example, $r$ is the geometric mean
of the semi-major and semi-minor axes of a compact object in an
inclined elliptical orbit. In an approximately circular orbit $r$ is
the orbital radius and for a particle moving faster than the escape
velocity $r$ is the impact parameter.

In the strong field regime where $r \sim M / m_{pl}^2$ the curvature
scale is also $\sim M / m_{pl}^2$ implying that $r/\cR \sim m / M$
whence  a perturbative expansion in $\ve$ is equivalent to one in
$m/M$.

The typical variation in time and space of the background is $\gsim
\cR$.
The wavelength $\lambda$ of radiated metric perturbations from the
compact object in a bound orbit is
\begin{eqnarray}
    \lambda \sim \sqrt{ \frac{ m_{pl}^2 r^3 }{ M } } \sim \cR ,
\end{eqnarray}
which shows that the wavelength of the gravitational waves does not
provide a separate scale independently from $\cR$.

\subsection{The CS-EFT approach: Issues and main features}

The effective field theory approach was first introduced for the
consideration of gravitational radiation from post-Newtonian binary systems in \cite{GoldbergerRothstein:PRD73}, spinning compact
objects in \cite{Porto:PRD73, PortoRothstein:PRL97,
Porto:grqc0701106, PortoRothstein:0712.2032}, and dissipative effects
due to the absorption of gravitational waves in
\cite{GoldbergerRothstein:PRD73_2, Porto:0710.5150}. See
\cite{Goldberger:LesHouches, PortoSturani:grqc0701105,
GoldbergerRothstein:GRG38} for introductory reviews. Let us call
these theories PN-EFT: they are constructed to describe the motion of
two slowly moving compact objects in a \emph{flat background}. In
particular, the compact objects are treated as effective \emph{point
particles}, the worldlines of which carry non-minimal operators
describing the multipole moments from companion-induced tidal
deformations as well as possible spin degrees of freedom and other
intrinsic moments. Below we describe briefly some general features of
EFT and the specific differences between our new CS-EFT approach and
the existing PN-EFT.

The use of point particles to source the metric perturbations  about
the flat background spacetime (note that the high frequency waves in
a quantum description corresponds to massless spin-two particles, the
gravitons, in flat space quantum field theory) prompts the appearance
of divergences. Fortunately there exists a well-established bank of
tools and techniques in quantum field theory for regularizing these
divergences and renormalizing the parameters and coupling constants
of the theory. A theory is renormalizable in the effective field
theory sense if observables are calculated in the low energy limit:
the divergences can always be absorbed into a renormalization of the
coupling constants of the infinite number of non-minimal worldline
operators. The use of dimensional regularization is particularly
useful in effective field theories because the renormalization group
equations are mass-independent for this scheme indicating that only
logarithmic divergences contribute to the renormalization procedure
\footnote{The effective field theory approach has also been used in \cite{Leibovich:NRGRtalk} to
derive the radiation reaction force on an electrically charged
extended body interacting with its own electromagnetic radiation,
generalizing the Abraham, Lorenz and Dirac (ALD) equation for a point
charge. It can also be derived from a
stochastic field theory perspective, which contains EFT. See
\cite{JohnsonHu:PRD65, JohnsonHu:FoundPhys35, Johnson:PhD}.}.

Our CS-EFT approach differs from this group of work in two ways.
First, we work with an arbitrary \emph{curved spacetime}. Second,
we allow for the compact object to move with \emph{relativistic
speeds} in \emph{strong field} regions of the background spacetime.
The post-Newtonian effective field theory of
\cite{GoldbergerRothstein:PRD73} treats bodies moving slowly through
a weak gravitational field.

\subsubsection{In-In formulation for real and causal equations of motion}

Technically there are also fundamental differences. To derive real
and causal equations of motion we emphasize the need to use the
closed-time-path  (CTP)  integral formalism based on an in-in
generating functional \cite{Schwinger:JMathPhys2, Keldysh:JEPT20,
ZhouSuHaoYu:PhysRep118, Jordan:PRD33, CalzettaHu:PRD35,
CalzettaHu:PRD37} (`in' and `out' here refer to the initial and final
vacua used to define the vacuum transition amplitude in the
generating functional). The authors of
\cite{GoldbergerRothstein:PRD73} use the in-out formalism which is
acceptable for field theories in a flat background spacetime since
the in- and out-vacua are equivalent up to an irrelevant phase in
that special case. In the presence of spacetime curvature, the
difference becomes serious, as we will show in \cite{Galley:EFT2} for
the EMRI scenario. The in-out formalism can calculate matrix elements
in scattering processes, but not expectation values of physical
observables in real-time evolution. The equations of motion for the
effective particle dynamics obtained from an in-out formulation are
not  generally  causal. An initial value formulation of quantum field
theory via the CTP generating functional is the only correct way for
the description of the system's evolution -- it guarantees real and
causal equations of motion for the particle dynamics
\cite{Jordan:PRD33, CalzettaHu:PRD37}.

\subsubsection{Effective point particle description}

Another  important
ingredient in our construction is an effective point particle
description for the motion of the compact object. Going beyond the
point particle approximation is necessary to include the effects of
tidal deformations induced by the background curvature as well as the
effects from spin and other intrinsic moments.  Following
\cite{GoldbergerRothstein:PRD73},  we introduce all possible terms
into the point particle action that are consistent with general
coordinate invariance and reparameterization invariance (and
invariance under $SO(3)$ rotations for a non-spinning spherically
symmetric compact object). By
implementing a matching procedure using coordinate invariant
observables we can match the observables of the effective point
particle theory with the long wavelength limit of observables in the
full ``microscopic" theory to determine the values of the coupling
constants of the non-minimal terms. As we will show in Section
\ref{sec:finsize} this allows us to deduce the order at which finite
size effects affect the motion of the compact object through the
statement of an \emph{Effacement Principle}. To our knowledge this
has not been explicitly given in the literature before for the EMRI scenario.

\subsubsection{The power counting rules}

Power counting is a generalization of dimensional analysis. In our
perturbative treatment it is crucial for determining how the Feynman
rules scale with the parameter $\ve$.  Once the scaling of the
Feynman rules are known we determine all of the tree-level Feynman
diagrams that appear at a particular order. Those diagrams containing
graviton loops are safely ignored. We also assemble the diagrams that
include the non-minimal worldline operators describing the finite
size of the compact object. Significantly, this allows us to
determine the order in $\ve$ at which finite size effects enter the
particle equations of motion. With the power counting rules  the EFT
approach becomes an efficient and systematic framework for
calculating the self-force to any order in perturbation theory.
Furthermore, by knowing how each Feynman diagram scales with $\ve$ we
can study a particular physical interaction that is of interest by
focusing our attention on a single diagram or on a few diagrams
without having to calculate every contribution that appears at that
order and at lower orders. For example, the leading order spin-spin
interaction (spin here refers to that of the compact object, not of
the SMBH) contributes to the self-force at third order in $\ve$ for a
maximally rotating compact object  and can be calculated from the
appropriate Feynman diagram \cite{Galley:PhD, Galley:spin}. The power counting
rules, the Feynman rules and their scaling with $\ve$ are derived in
Sections \ref{ch3:powercounting} and \ref{ch3:Feynmanrules}.

\subsubsection{Divergences and Regularization}

In Section \ref{ch3:reg} we propose a method for regularizing the
divergences that appear in the effective action. Our approach
utilizes a mixture of distributional and momentum space techniques
within the context of dimensional regularization. We know from
previous work 
 that the finite part of the self-force is
generally non-local and history dependent. However, the ultraviolet
divergences are quasi-local and independent of the history of the
effective point particle's motion. To isolate the quasi-local
divergence from the non-local finite part we use the method of
Hadamard's {\it  partie finie}, or finite part, from distribution
theory. (See Appendix \ref{app:distro} for a brief review of
the definitions and concepts of distribution theory relevant in this
work.)

After isolating the divergence from the non-local, finite remainder
we then use a momentum space representation for the propagator  in a
curved background, first derived for a scalar field by Bunch and
Parker \cite{BunchParker:PRD20},  to calculate the divergent
contributions. 
Their method is straightforward but not efficient for higher spin fields,
including 
gravitons in a curved space. (See also the related work of \cite{HuOConnor:PRD30}.)

A novel method applicable for any tensor field is developed in
\cite{Galley:momentum} for computing the momentum space
representation of the Feynman propagator. The method is sufficiently
general to do the same for any quantum two-point function, including
the retarded propagator $D^{ret}_{\alpha \beta \gamma^\prime
\delta^\prime} (x,x^\prime)$.  This approach makes use of
diagrammatic techniques borrowed from perturbative quantum field
theory. In Riemann normal coordinates, we expand the field action in
terms of the displacement from the point $x$. The series can be
represented in terms of Feynman diagrams, which allows for an
efficient evaluation of each term in the expansion. Furthermore, we
prove that some of the diagrams are zero to all orders. This identity
is not recognized in \cite{BunchParker:PRD20} even though its
relation to certain steps made in their calculations is evident.

\subsection{MST-QW Equation}

Assembling all these essential ingredients, in Section
\ref{sec:EFTMST-QW} we give a demonstration of how the curved
spacetime effective field theory construction is implemented, outline
the steps in the regularization of divergences, perform an actual
calculation of the effective action and from it derive the equation
of motion for the compact object including the effect of
gravitational self-force. As an application we work to first order in
the mass ratio  (i.e., $\ve$) and obtain the well-known
Mino-Sasaki-Tanaka-Quinn-Wald (MST-QW) equation
\cite{MinoSasakiTanaka:PRD55, QuinnWald:PRD56}.  This also sets the
stage for calculating the second order self-force,  the emitted
gravitational waves  and the motion of compact objects with spin
\cite{Galley:EFT2, Galley:EFT3, Galley:spin}.

\section{Effective field theory for post-Newtonian binaries and EMRIs}
\label{ch3:eft4grav}

Before proceeding to construct a curved spacetime effective field
theory for extreme mass ratio inspirals we briefly summarize the
original work of \cite{GoldbergerRothstein:PRD73}, which introduces
effective field theory techniques for describing post-Newtonian
binary sources of gravitational radiation.

\subsection{EFT in flat space for post-Newtonian binaries}

The aim of \cite{GoldbergerRothstein:PRD73} is to describe the motion of two slowly moving bodies through a weak gravitational field using effective field theory techniques in order to generate a perturbative expansion in powers of the relative velocity. One of the many benefits of using an effective field theory approach is that the method is systematic and efficient so that there is in principle no obstacle calculating to any order in the velocity.

The authors in \cite{GoldbergerRothstein:PRD73} start by  replacing
the compact objects with effective point particles. These are
described by an action consisting of the usual point particle action
plus all possible self-interaction terms that are consistent with
general coordinate invariance and reparameterization invariance of
the worldline. Then, the in-out generating functional is introduced
to derive an effective action
\begin{eqnarray}
    && {\hskip-0.4in} \exp \Big\{ i S_{eff} [z] \Big\} \nonumber \\
    && \equiv \int {\cal D} h_{\mu\nu} \exp \Bigg\{ i S_{pp} [z, \eta+ h] + i S [ \eta+ h ] \Bigg\}
\end{eqnarray}
where $S_{pp}$  is the effective point particle action, $S[\eta+
h]$ is the Einstein-Hilbert action for the full spacetime metric and $z^\alpha$ are the coordinates of the particle worldline.

Before integrating out the metric perturbations  the authors observe
that it is useful to separate $h_{\mu\nu}$ into potential
$H_{\mu\nu}$ and radiation $\bar{h}_{\mu\nu}$ contributions
\begin{eqnarray}
    h_{\mu\nu} = H_{\mu\nu} + \bar{h}_{\mu\nu}  .
\end{eqnarray}
This is suggested by the fact that the slowly moving bodies see a
nearly instantaneous gravitational potential and yet radiate
gravitational waves due to their relative accelerations. This
decomposition is also required to make the Feynman diagrams all scale
homogeneously with the relative velocity, $v$. In this way, the
perturbative expansion in $v$ is consistent and can be constructed to
any order.

Integrating out the potential gravitons using perturbation theory yields a theory of point particles moving in potentials. The radiation gravitons and the particle worldlines are non-dynamical at this stage and can be treated as external sources. In this effective theory, valid at the orbital scale of the binary, the authors derive the Einstein-Infeld-Hoffman potential \cite{EinsteinInfeldHoffmann:AnnMath39} as a check of their method.

The last effective theory the authors have constructed involves
integrating out the radiation gravitons. 
They then derive the famous power spectrum for quadrupolar
gravitational radiation by calculating the first non-vanishing
contribution to the imaginary part of the effective action; the real
part of the effective action generates equations of motion while the
imaginary part is related to the power of the emitted gravitational
radiation. 

Using an effective field theory approach it is not too surprising
that some of the parameters of the theory undergo classical
renormalization group (RG) scaling. In fact, the appearance of such
RG scaling is used by the authors to show that there are no finite
size effects up to $v^6$ order. In their words, ``whenever one
encounters a log divergent integral at order $v^6$ in the potential,
one may simply set it to zero. Its value cannot affect physical
predictions." \cite{GoldbergerRothstein:PRD73} This is how they
resolve the problem of the undetermined regularization parameters
that appear, at third post-Newtonian order, from regularizing the
singular integrals encountered with the traditional PN expansion
techniques. See \cite{Blanchet:LRR} and references therein for a complete discussion of the regularization ambiguity.


\subsection{EFT in curved spacetime for EMRIs}

Our construction of an EFT does not rely on the slow motion of the
bodies nor on the assumption that they move through a weakly curved
region of spacetime. Quite the contrary, we allow for the compact
object to move relativistically through the strong field region of
the SMBH background spacetime. As a result, the metric perturbations
generated by the motion of the compact object cannot be partitioned
simply into an instantaneous potential and radiation contributions.


Within LISA's observable frequency bandwidth, $\ve$ takes values between
$\sim 10^{-5}$ and $10^{-7}$. Being so small over the
dynamical time scale of the inspiral implies that $\ve$ is a good parameter for
building a perturbation theory within the context of effective field
theory.

Utilizing the dissimilar magnitudes of the compact object's size and
the background curvature scale, we can construct two kinds of
effective field theories. The first describes the compact object, in
isolation from other external sources, as an effective point
particle. By allowing for all possible worldline self-interaction
terms that are consistent with the symmetries of the theory we can
account for the tidal deformations, spin and intrinsic multipole
moments that the compact object may experience when it does interact
with external sources. The second EFT is valid at scales $ \gsim \cR$
and results from integrating out the metric perturbations. The
resulting theory is that of an effective point particle subjected to
a self-force from the gravitational radiation reaction as a result of
its motion in the background spacetime, with the force and the
radiation  evolving self-consistently. Using a matching procedure we
can establish the values of the coupling constants appearing in the
effective point particle action.

\subsection{EFT of a compact object} \label{ch3:effpp}

In applying the EFT formalism we first
construct an effective point particle theory for the small mass $m$.
This allows for a point particle description of the compact object's
motion through the background spacetime while taking into account any
tidally induced moments, or finite size effects, that might affect
its motion. 
An example is provided in classical electromagnetism wherein long-wavelength radiation scatters off a small, metallic sphere possessing no residual charge. The interaction of the field induces a time-varying dipole moment on the surface of the sphere. Since the wavelength is much larger than the sphere, this system can effectively be described by radiation interacting with a point particle carrying an induced dipole moment on its worldline.

In the full theory describing the motion of a neutron star and the dynamics of the spacetime metric it moves in, the total action is given by
\begin{eqnarray}
    S_{tot} = S [g] + S_{ns} [g; \rho, p \ldots]  .
\end{eqnarray}
The quantities in the neutron star action, $\rho, p, \ldots$, are the appropriate hydrodynamic variables necessary to describe the internal dynamics of the neutron star whatever its equation of state. If the compact object under consideration is a small black hole then there is only the dynamics of vacuum spacetime to consider, which is described entirely by the Einstein-Hilbert action
\begin{eqnarray}
    S [g] = 2 m_{pl}^2 \int d^4x \, g^{1/2} R
\end{eqnarray}
where $R$ is the Ricci curvature scalar of the spacetime and $g$ is the absolute value of the metric's determinant.

The effective point particle description of the compact object is constructed by ``integrating out" the short wavelength gravitational perturbations up to the scale $r_m \sim m / m_{pl}^2$. In doing so we introduce an effective point particle action $S_{pp}$ to describe the motion of the compact object moving in the combined geometry of the background SMBH plus long wavelength gravitational perturbations so that the total action becomes
\begin{eqnarray}
    S_{tot} = S [g] + S_{pp} [z,g]  .  \label{totaction0}
\end{eqnarray}
Here $z^\alpha (\lambda)$ are the coordinates of the particle worldline and $\lambda$ is its affine parameterization.

Being a description of the extended compact object the effective
point particle action should include all possible self-interaction
terms that are consistent with the symmetries of the theory \cite{Burgess:EFT}, which
are general coordinate invariance and worldline reparameterization
invariance. For the discussion here, we will assume that the compact
object is perfectly spherical when removed from external influences (e.g. background curvature) so that it carries no permanent moments. This implies, for example, excluding spinning compact
objects in our construction, at least for now \footnote{We will
introduce spin and determine the influence it has on the effective
particle's motion in a forthcoming paper \cite{Galley:spin}.}. Hence,
$S_{pp}$ should also be invariant under $SO(3)$ rotations. From
these considerations the most general such action is
\begin{eqnarray}
    S_{pp} [ z, g] &=& - m \int d\tau + c_R \! \int \! d\tau \, R + c_V \! \int \! d\tau \, R_{\mu\nu} \dot{z}^\mu \dot{z}^\nu \nonumber \\
    && + c_E \! \int \! d\tau \, \cE_{\mu\nu} \cE^{\mu\nu}  + c_B \! \int \! d\tau \, \cB_{\mu\nu} \cB^{\mu\nu} + \cdots ,  \nonumber \\
    && \label{effpp0}
\end{eqnarray}
which is effectively an expansion in powers of the compact body's radius $r_m$ over the much longer wavelength of the gravitational waves $\lambda$. This can be interpreted as a multipole expansion where the multipoles carry information about the induced moments that the background curvature and long wavelength gravitational perturbations impart to the compact object. We showed earlier that in the strong field region of a SMBH the wavelength of the metric perturbations is the same order as the curvature scale of the background spacetime, $\lambda \sim \cR$. This implies that the terms in the above multipole expansion have a definite scaling with $\ve$, which we will later demonstrate in Section \ref{sec:finsize}.

The symmetric and traceless tensors $\cE_{\mu\nu}$ and $\cB_{\mu\nu}$ are the electric and magnetic parts of the Weyl curvature, defined as
\begin{eqnarray}
    \cE_{\mu\nu} &=& C_{\mu\alpha \nu \beta} \dz^\alpha \dz^\beta \\
    \cB_{\mu\nu} &=& \epsilon_{\mu \alpha \beta \lambda} C^{\alpha \beta} _{~~~\nu \rho} \dz^\lambda \dz^\rho
\end{eqnarray}
where $\dz^\alpha$ is the particle's 4-velocity. Recall that these tensors are orthogonal to $\dz^\alpha$ and $C_{\mu\alpha \nu \beta} = R_{\mu \alpha \nu \beta}$ for a Ricci flat spacetime.

The terms in the effective point particle action (\ref{effpp0}) proportional to the Ricci curvature vanish at leading order in $\ve$ 
. The equations of motion for the full metric
\begin{eqnarray}
    R_{\mu\nu} - \frac{1}{2} g_{\mu\nu} R = T_{\mu\nu} ^{pp} (z,g) = O(\ve)  ,
\end{eqnarray}
where $T_{\mu\nu} ^{pp}$ is the stress-energy tensor of the effective point particle, can be used to set the second and third terms in (\ref{effpp0}) to zero at leading order. Equivalently, for the term proportional to $c_R$, for example, we can redefine the metric $g_{\mu\nu}$ in terms of a new metric $g_{\mu\nu}^\prime$ through the field redefinition \cite{GoldbergerRothstein:PRD73}
\begin{eqnarray}
    g_{\mu\nu} (x) = g^\prime_{\mu\nu} (x) \left[ 1 + \frac{ \xi }{ 2 m_{pl}^2 } \int d\tau \frac{ \delta^4 (x - z(\tau) ) }{ g^{\prime 1/2} } \right]   .   \label{fieldredef1}
\end{eqnarray}
This transformation implies that the Einstein-Hilbert action is, to linear order in the (small) arbitrary parameter $\xi$,
\begin{eqnarray}
    && {\hskip-0.4in} 2 m_{pl}^2 \int d^4x \, g^{1/2} R (g) \nonumber \\
    && = 2 m_{pl}^2 \int d^4x \, g^{\prime 1/2} R(g^\prime) + \xi \int d\tau \, R  .
\end{eqnarray}
The term in $S_{pp}$ linear in $R$ then appears with the constant $c_R+\xi$, which can be set to zero since $\xi$ is arbitrary. One can find a similar field redefinition to remove the term proportional to $c_V$.
(See \cite{Politzer:NuclPhysB172, Georgi:NuclPhysB361} for a more thorough discussion of removing such terms from the effective field theory.)
Using the metric field equations or, equivalently, performing a field redefinition of the metric one can remove all occurrences of the Ricci tensor in the effective point particle action. It follows that  the non-minimal couplings in $S_{pp}$ contain terms that depend only on the Riemann curvature tensor.

These field redefinitions allow for the effective point particle action (\ref{effpp0}) to be written as
\begin{eqnarray}
    S_{pp} [z, g] &=& - m \int d\tau + c_E \int d\tau \, \cE_{\mu\nu} \cE^{\mu\nu} \nonumber \\
    && + c_B \int d\tau \, \cB_{\mu\nu} \cB^{\mu\nu}  + \cdots .  \label{effpp1}
\end{eqnarray}
In a later section we will show that the non-minimal couplings in
$S_{pp}$ are entirely negligible for calculating the MST-QW (first
order) self-force. The MST-QW self-force is sufficient for computing
gravitational waveforms and generating  templates for LISA to {\it
detect} gravitational waves from EMRIs \footnote{Actually, this is
more than sufficient as recent work suggests that the less accurate
``kludge" waveforms may be adequate for the detection phase of at
least a certain class of EMRIs \cite{GlampedakisBabak:CQG23,
Drasco:CQG23}.}.
However, a more precise determination of the masses, spins, etc. of the binary constituents requires more accurate
higher order templates, which can be computed by knowing the higher
order contributions to the self-force \cite{Burko:PRD67, Rosenthal:CQG22}.

In the next section we derive the first order equations of motion for
the compact object using the CS-EFT approach. These equations, which
describe the self-force on the mass $m$, were previously found by
Mino, Sasaki and Tanaka \cite{MinoSasakiTanaka:PRD55} using matched
asymptotic expansions and independently by Quinn and Wald
\cite{QuinnWald:PRD56} using axiomatic methods. In principle, we can
compute the formal equations of motion to higher orders in $\ve$
thereby extending the work of \cite{MinoSasakiTanaka:PRD55} and
\cite{QuinnWald:PRD56}. The second paper in this series
\cite{Galley:EFT2} will give results through the second order in
$\ve$ .

\section{CS-EFT derivation of MST-QW  Equation for first order self-force}
\label{sec:EFTMST-QW}

In the previous section, we outlined the construction of an effective field theory that replaces the extended compact object by an effective point particle. 
In this section we construct  an EFT for the motion of the effective
particle by integrating out the metric perturbations at the scale of
the radius of curvature $\cR$. In doing so, we derive the MST-QW
self-force on the compact object.


Denote the background (unperturbed) metric by $g_{\mu\nu}$ so that
the  total metric is given by the background plus the perturbations
generated by the presence  of the moving compact object
\begin{eqnarray}
    g^{\rm full}_{\mu\nu} = g _{\mu\nu} + \frac{ h_{\mu\nu} }{m_{pl} }  .
\end{eqnarray}
The metric perturbations $h_{\mu\nu}$ are presumed to be small so
that $| h_{\mu\nu} | \ll m_{pl}$. We will occasionally make use of
the shorthand notation
\begin{eqnarray}
    \bar{h}_{\mu\nu} \equiv \frac{ h_{\mu\nu} }{ m_{pl} }
\end{eqnarray}
for the (dimensionless) ratio of the metric perturbation to the
Planck mass. From (\ref{totaction0}) the total action describing the
interactions between the metric perturbations and the particle is
given by the sum of the Einstein-Hilbert and effective point particle
actions,
\begin{eqnarray}
    S_{tot} [ g + \barh, z] = S [g + \barh] + S_{pp} [ g + \barh, z]  .
\end{eqnarray}
We expand the Einstein-Hilbert action in orders of $h_{\mu\nu}$
\begin{eqnarray}
    S [ g + \barh] &=&  \frac{1}{2} \int d^4x \, g^{1/2} \Bigg[ 2 h_{\alpha \beta ; \gamma} h^{\alpha \gamma ; \beta} - h_{\alpha \beta; \gamma} h^{\alpha \beta ; \gamma} \nonumber \\
    && {\hskip0.5in} - 2 h_{;\alpha} \Big( h^{\alpha \beta } _{~~;\beta} - \frac{1}{2} \, h^{;\alpha} \Big) \Bigg] \nonumber \\
    &+& \sum_{n=3}^\infty \frac{1}{n!} \int d^4x \, g^{1/2} V_{ (n) } ^{\alpha_1 \cdots \alpha_{2n} \rho_1 \rho_2}  \nonumber \\
    && ~~ \times \nabla_{\rho_1} h_{\alpha_1 \alpha_2} \nabla_{\rho_2} h_{\alpha_3 \alpha_4}  h_{\alpha_5 \alpha_6} \cdots h_{\alpha_{2n-1} \alpha_{2n} } \nonumber \\
    &\equiv& S^{(2)} + S^{(3)} + \cdots,     \label{expandEH1}
\end{eqnarray}
where $S^{(n)}$ denotes the part of the action containing terms
proportional to $n$ factors of $h_{\mu\nu}$. The quadratic
contribution is the kinetic term for $h_{\mu\nu}$ and provides the
propagator corresponding to some appropriate boundary conditions
(e.g., retarded, Feynman).
The action is invariant under infinitesimal coordinate transformations on the background spacetime.

We also need to expand the point particle action in powers of
$h_{\mu\nu}$. Using (\ref{effpp0}) we find
\begin{eqnarray}
    && {\hskip-0.5in} S_{pp} [ z, g +\barh] \nonumber \\
    &=& -m \int d\lambda \left( - g_{\alpha \beta} \dz^\a \dz^\beta \right)^{1/2} \nonumber \\
    && + \sum_{n=1}^\infty \frac{1}{n!} \int d\lambda \, V_{pp \, (n) }^{ \alpha_1 \cdots \alpha_{2n} }  h_{\alpha_1 \alpha_2} \cdots h_{\alpha_{2n-1} \alpha_{2n} } \nonumber \\
    &\equiv& S_{pp}^{(0)} + S_{pp} ^{(1)} + S_{pp}^{(2)} + \cdots
\end{eqnarray}
where the first two integration kernels are
\begin{eqnarray}
    V_{pp \, (1)} ^{ \alpha \beta} &=& \frac{ m }{2m_{pl} } \frac{ \dz^\alpha \dz^\beta}{ \big( - g_{\mu\nu} \dz^\mu \dz^\nu \big) ^{1/2} } \\
    V_{pp \, (2)}^{ \alpha \beta \gamma \delta} &=& - \frac{m}{4 m_{pl}^2} \frac{ \dz^\alpha \dz^\beta \dz^\gamma \dz^\delta }{ \big( - g_{\mu\nu} \dz^\mu \dz^\nu \big) ^{3/2} }
\end{eqnarray}
and where $S_{pp}^{(n)}$ denotes the part of the action for the
point particle containing terms proportional to $n$ factors of $h_{\mu\nu}$.


\subsection{The closed-time-path effective action}
\label{ch3:ctp}

The construction of an effective field theory for the motion of the effective point particle in a curved spacetime begins with the CTP, or in-in, generating functional
\begin{eqnarray}
    && {\hskip-0.4in} Z [ j^\mu, j ^{\prime \mu}, J^{\mu\nu} , J^{\prime \mu\nu} ]  \nonumber \\
    &=&  \int _{CTP} {\hskip-0.15in} {\cal D} z^\mu \int_{CTP} {\hskip-0.15in} {\cal D} h^{\mu\nu} \, \exp \Bigg\{ i S[ g + \barh ] - i S [ g +\barh^\prime]  \nonumber \\
    && {\hskip0.3in} + i S_{pp} [z, g + \barh] - i S_{pp} [z^\prime, g + \barh^\prime]  \nonumber \\
    && {\hskip0.3in} + i \int d\lambda ( j_\mu z^\mu - j_{\mu}^\prime z^{\prime \mu} )  \nonumber \\
    && {\hskip0.3in} + i  \int d^4x \, g^{1/2} ( J_{\mu\nu} h^{\mu\nu} - J^{\prime} _{\mu\nu} h^{\prime \mu\nu} )  \Bigg\}  ,    \label{genfun1}
\end{eqnarray}
where
\begin{eqnarray}
    && {\hskip-0.5in} \int_{CTP} {\cal D} z^\mu \, ( . )  \nonumber \\
    &\equiv& \int dz^\mu_f \int dz_i ^\mu \int dz_i^{\prime \mu} \int_{z_i^\mu} ^{z_f^\mu} {\cal D}z^\mu \int _{z_i^{\prime \mu} } ^{z_f^\mu} {\cal D} z^{\prime \mu} \, (.) ~~
\end{eqnarray}
and likewise for the graviton CTP path integral.


We find it more convenient to relabel the unprimed and primed
variables with a lowercase Latin index $a,b,c,...$ (from the
beginning of the alphabet): $1$ for an unprimed index and $2$ for a
primed index, respectively. We introduce the so-called CTP metric
$c_{ab}$ that lowers and raises these indices where
\begin{eqnarray}
    c_{ab} = \left(
        \begin{array}{cc}
            1 & 0 \\
            0 & -1
        \end{array} \right) = c^{ab}  .  \label{ctpmetric0}
\end{eqnarray}
For a current ``contracted" with a scalar field, for example, the notation implies
\begin{eqnarray}
    J^a \Phi_a &\equiv& c^{ab} J_a  \Phi_b \\
    &=& J_1 \Phi_1 - J_2 \Phi_2 = J^1 \Phi^1-J^2 \Phi^2 \\
    &=& J \Phi - J^\prime \Phi^\prime  .
\end{eqnarray}
The Einstein-Hilbert action can be written as
\begin{eqnarray}
    S [g + \barh_a] &\equiv& S [ g + \barh_1 ] - S [g + \barh_2 ] \nonumber \\
    &=& S [ g + \barh ] - S [g + \barh^\prime]  .
\end{eqnarray}
and similiarly for $S_{pp}$. A comprehensive description of the CTP formalism using these notations can be found in \cite{RamseyHu:PRD56, JohnsonHu:PRD65}. In this compact notation the CTP generating functional takes on the form,
\begin{eqnarray}
    && {\hskip-0.15in} Z [ j^\mu_a  , J^{\mu\nu} _a ]  \nonumber \\
    && = \int _{CTP} {\hskip-0.15in} {\cal D} z^\mu \int_{CTP} {\hskip-0.15in} {\cal D} h^{\mu\nu} \, \exp \Bigg\{ i S[ g + \barh_a ] + i S_{pp} [z_a, g + \barh_a]  \nonumber \\
    && {\hskip0.2in} + i \int d\lambda \, j^a_\mu z^\mu_a +  i  \int d^4x \, g^{1/2}  J^a_{\mu\nu} h_a^{\mu\nu}  \Bigg\}  .   \label{genfun2}
\end{eqnarray}
Notice the similarity in appearance to the in-out generating
functional yet the substantive  difference in physical consequences mentioned earlier.

Calculating derivatives of the generating functional with respect to the external current $J^a_{\mu\nu}$  generates time-ordered correlation functions of the quantum metric perturbations $\hath_a^{\mu\nu}$
\begin{eqnarray}
   && {\hskip-0.5in} \big\langle 0, {\rm in} \big| \bar{T} \, \hat{h}_{a_1}^{\mu_1\nu_1} (x_1) \cdots \hat{h}_{a_n} ^{\mu_n \nu_n} (x_n) \big| 0, {\rm in } \big\rangle \nonumber \\
    && = \frac{1}{Z} \, \frac{ \delta^n Z}{ \delta i J^{a_1} _{\mu_1 \nu_1} (x_1) \cdots \delta i J^{a_n} _{\mu_n \nu_n } } .  \label{inincorr1}
\end{eqnarray}
The CTP time-ordering operator $\bar{T}$ is defined so that unprimed operators are time-ordered, primed operators are anti-time-ordered and all primed operators are ordered to the left of unprimed operators. For example, the full graviton Feynman propagator is calculated from the generating functional by
\begin{eqnarray}
    i D_F^{\alpha \beta\gamma \delta} (x,x^\prime) &=& \big\langle 0, {\rm in} \big| T \, \hat{h}_1^{\alpha \beta} (x) \hat{h}_1^{\gamma \delta} (x^\prime) \big| 0, {\rm in} \big\rangle \nonumber \\
    &=& \left. \frac{1}{Z} \, \frac{ \delta^2 Z }{ \delta i J^1_{\alpha \beta} (x) \, \delta i J^1_{\gamma \delta} (x^\prime) } \right|_{ j^a_\mu, J^a_{\mu\nu} = 0 }  .  \nonumber \\
    && \label{fullpropagator}
\end{eqnarray}
Notice that (\ref{inincorr1}) describes the full graviton correlation functions and includes the effects from (nonlinear) particle-field interactions. Likewise, derivatives of the generating functional with respect to $j^a_\mu$ generate time-ordered correlation functions of the worldline coordinates including the back-reaction from the metric perturbations.

To guarantee a well-defined graviton propagator on the background spacetime we adopt the
Faddeev-Popov \cite{FaddeevPopov:PhysLett25} gauge-fixing procedure by introducing the action
\begin{eqnarray}
    S_{gf} = - m_{pl}^2 \int d^4x \, g^{1/2} G_\alpha G^\alpha  ,
\end{eqnarray}
which is equivalent to picking the gauge $G_\alpha [ h_{\mu \nu} ]
\approx 0$ for the metric perturbations. (The $\approx$ denotes weak
equality in the sense of Dirac \cite{Dirac}.) Since we will be
dealing with tree-level interactions only there is no need to
introduce ghost fields into the gravitational action.

We choose the Lorenz gauge for the trace-reversed metric perturbations, defined as
\begin{eqnarray}
    \psi_{\alpha \beta} \equiv h_{\alpha \beta} - \frac{1}{2} g_{\alpha \beta} h  ,
\end{eqnarray}
so that the gauge-fixing function is
\begin{eqnarray}
    G_\alpha [h_{\mu\nu} ] = \psi_{\alpha \beta} ^{~~~;\beta} = h_{\alpha \beta} ^{~~~;\beta} - \frac{1}{2} h_{;\alpha}  \approx 0  .
\end{eqnarray}
In this gauge, the kinetic term in (\ref{expandEH1}) is
\begin{eqnarray}
    S^{(2)} &=& - \frac{1}{2} \int d^4x \, g^{1/2} \Bigg( h_{\alpha \beta; \gamma} h^{\alpha \beta ; \gamma} - \frac{ 1}{2} h_{;\alpha} h^{;\alpha}  \nonumber \\
    && {\hskip0.5in} - 2 h^{\alpha \beta} R_{\alpha ~\beta} ^{~\gamma ~\delta} h_{\gamma \delta} \Bigg)   ,   \label{mpkinetic1}
\end{eqnarray}
which applies to both the $h_1^{\alpha \beta}$ and $h_2^{\alpha \beta}$ metric perturbations (equivalently, the unprimed and primed fields, respectively).

The generating functional can now be written as
\begin{eqnarray}
    && {\hskip-0.25in}  Z [j_a^\mu, J_a^{\mu\nu} ] \nonumber \\
    &=&  \int _{CTP} {\hskip-0.15in} {\cal D}z^\mu \, \exp \Bigg\{ i S_{pp}^{(0)} [z_a] + i \int d\lambda \, j_\mu^a z_a^\mu \Bigg\}   \nonumber \\
    &&  \times \int_{CTP} {\hskip-0.15in} {\cal D} h^{\mu\nu} \, \exp \Bigg\{  i S^{(2)} [ h_a ]  + i \int d^4x \, g^{1/2} J_{\mu\nu} ^a h^{\mu\nu} _a   \nonumber \\
    && {\hskip0.4in} + \sum_{n=3}^\infty  i S^{(n)} [h_a ] + \sum_{n=1}^\infty i S_{pp}^{(n)} [z_a, h_a]   \Bigg\}  . \label{interactions1}
\end{eqnarray}
We have factored out the lowest order point particle contribution
from the graviton path integral  since it is independent of the
metric perturbations. This is the limiting situation when the
particle is regarded as a test object which  produces no
perturbations about the background metric (and thus its motion
follows  a geodesic of the background geometry.)

Perturbation theory in the in-in formalism is formulated similar to
that in the in-out formalism. The fact that the metric perturbations
couple linearly to the external current $J^{\mu\nu}_a$ implies that
every occurrence of the field in (\ref{interactions1}) can be
replaced by a functional derivative of the external current,
\begin{eqnarray}
    h^{\mu\nu}_a (x) \to -i \frac{ \delta }{ \delta J_{\mu\nu}^a (x) }  .
\end{eqnarray}
This rule allows for the generating functional to be written as
\begin{eqnarray}
    Z [ j^\mu_a, J_a^{\mu\nu} ] &=&  \int _{CTP} {\hskip-0.15in} {\cal D}z^\mu \, \exp \Bigg\{ i S_{pp} ^{(0)} [z_a] + i \int d\lambda \, j_\mu^a z^\mu_a \nonumber \\
    && + \sum_{n=1}^\infty i S_{pp}^{(n)} \left[ z_a, -i \frac{ \delta }{ \delta J^a_{\alpha \beta} } \right] \nonumber \\
    && + \sum_{n=3}^\infty  i S^{(n)} \left[  -i \frac{ \delta }{ \delta J^a_{\alpha \beta} } \right]  \Bigg\} Z_0 [ J_a ^{\mu\nu} ]  \label{genfun3}
\end{eqnarray}
since the interaction terms can now be taken outside of the graviton path integral. The quantity $Z_0$ is the free field generating functional for the metric perturbations
\begin{eqnarray}
    Z_0 [ J_a ^{\mu\nu} ] &\!\!=\!\!& \! \int_{CTP} {\hskip-0.15in} {\cal D} h^{\mu\nu} \, \exp \Bigg\{ i S^{(2)} + i \! \int \! d^4x \, g^{1/2} J_{\mu\nu} ^a h^{\mu\nu}_a \Bigg\}  \nonumber \\
    &&
\end{eqnarray}
and is calculated by integrating the Gaussian giving
\begin{eqnarray}
    Z_0 [ J_a ^{\mu\nu} ] = \exp \Bigg\{ - \frac{1}{2} \, J_a ^{\alpha \beta} \cdot D_{\alpha \beta \gamma^\prime \delta^\prime} ^{ab} \cdot J_b^{\gamma^\prime \delta^\prime} \Bigg\}
\end{eqnarray}
where the free graviton two-point functions are
\begin{eqnarray}
    D^{ab} _{\alpha \beta \gamma^\prime \delta^\prime} (x,x^\prime) = \left( \begin{array}{cc}
        D^F _{\alpha \beta \gamma^\prime \delta^\prime} & - D^- _{\alpha \beta \gamma^\prime \delta^\prime} \\
        - D^+ _{\alpha \beta \gamma^\prime \delta^\prime} & D^D _{\alpha \beta \gamma^\prime \delta^\prime}
        \end{array} \right)
\end{eqnarray}
(momentarily leaving out the CTP tensor indices, also
$D_{cd} = c_{ca} c_{db} D^{ab}$). Specifically, these are the Feynman
($D^F$) and Dyson propagators ($D^D$) and the positive/negative
frequency Wightman functions ($D^{\pm}$). See Appendix
\ref{app:twoptfns} for their definitions, identities and useful relations.

Upon defining the interaction Lagrangian as
\begin{eqnarray}
    \int d^4x \, {\cal L}_{int} \Bigg[ z_a, -i \frac{ \delta }{ \delta J^a_{\alpha \beta} } \Bigg] &=& \sum_{n=1}^\infty i S_{pp}^{(n)} \left[ z_a, -i \frac{ \delta }{ \delta J^a_{\alpha \beta} } \right] \nonumber \\
    && + \sum_{n=3}^\infty  i S^{(n)} \!\! \left[  -i \frac{ \delta }{ \delta J^a_{\alpha \beta} } \right] ~~~~~~~
\end{eqnarray}
we find that the generating functional can be written in the form
\begin{eqnarray}
    Z [ j^\mu_a, J^{\mu\nu}_a ] &\arreq&  \int_{CTP} {\hskip-0.1in} {\cal D}z^\mu \, \exp \Bigg\{ i S_{pp} ^{(0)} [z_a]  + i \int d\lambda \, j_\mu^a z_a^\mu \nonumber \\
    && + i \int d^4x \, {\cal L} _{int} \Bigg[ z_a, -i \frac{ \delta }{ \delta J_{\alpha \beta}^a} \Bigg] \Bigg\} Z_0 [J^a_{\mu\nu} ]  .  \nonumber \\
    &&     \label{genfun4}
\end{eqnarray}
Notice that this is expressed as a certain functional derivative operator acting on a Gaussian functional of the external currents $J^{\mu\nu}_a$.


Computing the Legendre transform of the generating functional with respect to the particle and graviton currents gives the effective action
\begin{eqnarray}
    \Gamma [ \langle \hat{z}^\mu_a \rangle, \langle \hat{h}^{\mu\nu} _a \rangle ] &=& - i \ln Z [ j^\mu_a, J^{\mu\nu}_a ] - \int d\lambda \, j_\mu^a \langle \hat{z}^\mu_a \rangle \nonumber \\
    && - \int d^4x \, g^{1/2} J_{\mu\nu}^a \langle \hat{h}^{\mu\nu}_a \rangle  .  \label{effaction0}
\end{eqnarray}
The equations of motion for the expectation values of the worldline coordinates and the metric perturbations are then found by varying the effective action,
\begin{eqnarray}
    && 0 = \frac{ \delta \Gamma }{ \delta \langle \hat{z}_a ^\mu \rangle } \Bigg| _{ z_1= z_2, h_1=h_2, j_a = J_a = 0 } \label{particleeom0} \\
    && 0 = \frac{ \delta \Gamma }{ \delta \langle \hat{h}_a ^{\mu\nu} \rangle } \Bigg| _{z_1= z_2, h_1=h_2,  j_a = J_a = 0}  \label{fieldeom0}
\end{eqnarray}
In this paper, we are interested in deriving the equations of motion
for the particle including the radiation reaction force (self-force).
We therefore focus our attention on (\ref{particleeom0}) and leave
the field equations describing the gravitational waves
(\ref{fieldeom0}) for a future paper \cite{Galley:EFT3}. This allows
us to set $J_a^{\mu\nu}=0$ in the remainder of this paper.

Before calculating the effective action we make a few remarks.
A classical equation of motion does not adequately describe the
particle's motion when the quantum mechanical fluctuations of the
worldline are not negligible. However, the particle's quantum
trajectories can be decohered by interactions with the quantum
fluctuations of the metric perturbations, or other matter fields
present, resulting in a classical worldline for the particle. We
call this the semiclassical limit (classical particle in a quantum
field). The condition for the existence of a semiclassical limit and
the appearance of stochastic behavior are stated in the beginning of
the Introduction section. See \cite{GalleyHu:PRD72,
GalleyHuLin:PRD74} and references therein for more details of issues
pertaining to a particle moving in an arbitrary curved background. So
as not to distract the reader with these issues it suffices to know
that for a solar mass compact object (treated as a point particle)
moving in the geometry of a $10^5 M_\odot$ Schwarzschild black hole that the
quantum mechanical fluctuations are of the order of $10^{-24}$cm
indicating that for EMRI astrophysical objects the semi-classical
limit is well-defined, as expected \cite{GalleyHuLin:PRD74}.

\subsection{Power counting rules}
\label{ch3:powercounting}

All of the terms following the  kinetic term $S^{(2)}$ in
(\ref{interactions1}) represent self interactions of the field and
various particle-field interactions. Each of these interaction terms
may be represented by a Feynman diagram.
To write down all of the relevant diagrams that contribute to the
effective action at a specific order in $\ve$ we need to
know how each of the interaction terms in (\ref{interactions1}) scale
with $\ve$ and $\cR$. The scaling behavior that we develop here are
called power counting rules and are essentially a generalization of
dimensional analysis. We first develop the power counting rules for
the parameters of the effective field theory; we ignore for now the
non-minimal point particle couplings in $S_{pp}$ (i.e., $c_E$, $c_B$, $\ldots$).

As discussed previously, the curvature scale $\cR$ describes the length scale of temporal and spatial variations of the curvature in the background geometry. This implies that each of the spacetime coordinates scale according to
\begin{eqnarray}
    x^\mu \sim \cR  . \label{coordscaling}
\end{eqnarray}
From the kinetic term for the metric perturbations we deduce that if $S^{(2)} \sim 1$ then
\begin{eqnarray}
    1 \sim \cR ^4 h^2 \left( \frac{1}{\cR} \right)^2 \sim \cR^2 h^2
\end{eqnarray}
and the metric perturbation  scales with $\cR$ as
\begin{eqnarray}
    h_{\mu\nu} \sim \frac{ 1}{ \cR}  .  \label{mpscaling}
\end{eqnarray}

The particle-field interactions, indicated by the terms $S_{pp}
^{(n)}$ for $n \ge 1$, contain inverse powers of the Planck mass,
$m_{pl}$. To power count these terms we form the ratio
\begin{eqnarray}
	\frac{ m}{m_{pl} } \sim \Big( \frac{ m }{ m_{pl}^2 \cR} \Big) \Big( \frac{ m_{pl} }{ m } \Big) \big( m \cR \big)  .
\end{eqnarray}
We remark that the factor $m \cR$ is the scale of the (conserved) angular momentum for a test particle following a geodesic in the strong-field region of the background spacetime since
\begin{eqnarray}
	L = m g_{\alpha \beta} \psi^\alpha \dx^\beta  \sim m \cR  \label{Lscaling}
\end{eqnarray}
where $\dx^\beta$ is the $4$-velocity of the geodesic. We therefore find that 
\begin{eqnarray}
    \frac{ m}{ m_{pl} } \sim \sqrt{ \ve L }   . \label{planckscaling}
\end{eqnarray}
(Neither does the quantity $m\cR$ vanish if a test particle plunges radially toward the SMBH (for which the angular momentum is zero) nor its value rely on the existence of any isometries of the background spacetime. In order to avoid problems of interpretation arising from scenarios where the particle's angular momentum vanishes one can identify $m\cR$ as the leading order classical particle action, which is non-zero for all geodesic motions; see (\ref{ppinterxns55}) below with $n=0$ \footnote{We thank T. Jacobson and A. Buonanno for bringing this subtlety to our attention.}. Regardless, we use the symbol $L$ to denote the quantity $m \cR$ in what follows.)

The four scaling laws in (\ref{coordscaling}), (\ref{mpscaling}), (\ref{Lscaling}) and (\ref{planckscaling}) determine the power counting rules for identifying the appropriate Feynman diagrams that enter into the evaluation of the effective action. See Table \ref{table:scaling}.

\begin{table}
    \centering
    \caption{Power counting rules}
    \label{table:scaling}
  \begin{center}
    \begin{tabular}{cccccccccccccccc}
        \hline
        \hline \\
        $x^\mu$ &~&~& $h_{\mu\nu}$ &~&~& $L$ &~&~& $\displaystyle{\frac{m}{m_{pl} } }$ &~&~& $S_{pp} ^{(n)}$ &~&~& $S^{(n)}$ \\ \\
        \hline \\
         $\cR$ &~&~& $\displaystyle{\frac{1}{\cR} }$ &~&~& $m \cR$ &~&~& $\sqrt{\ve L}$  &~&~& $\ve \displaystyle{\left( \frac{ L }{\ve} \right)^{1-n/2} }$ &~&~& $\displaystyle{\left( \frac{L}{\ve } \right) ^{1-n/2} }$ \\ \\
        \hline
    \end{tabular}
  \end{center}
\end{table}

\begin{figure}
    \center
    \includegraphics[width=8cm]{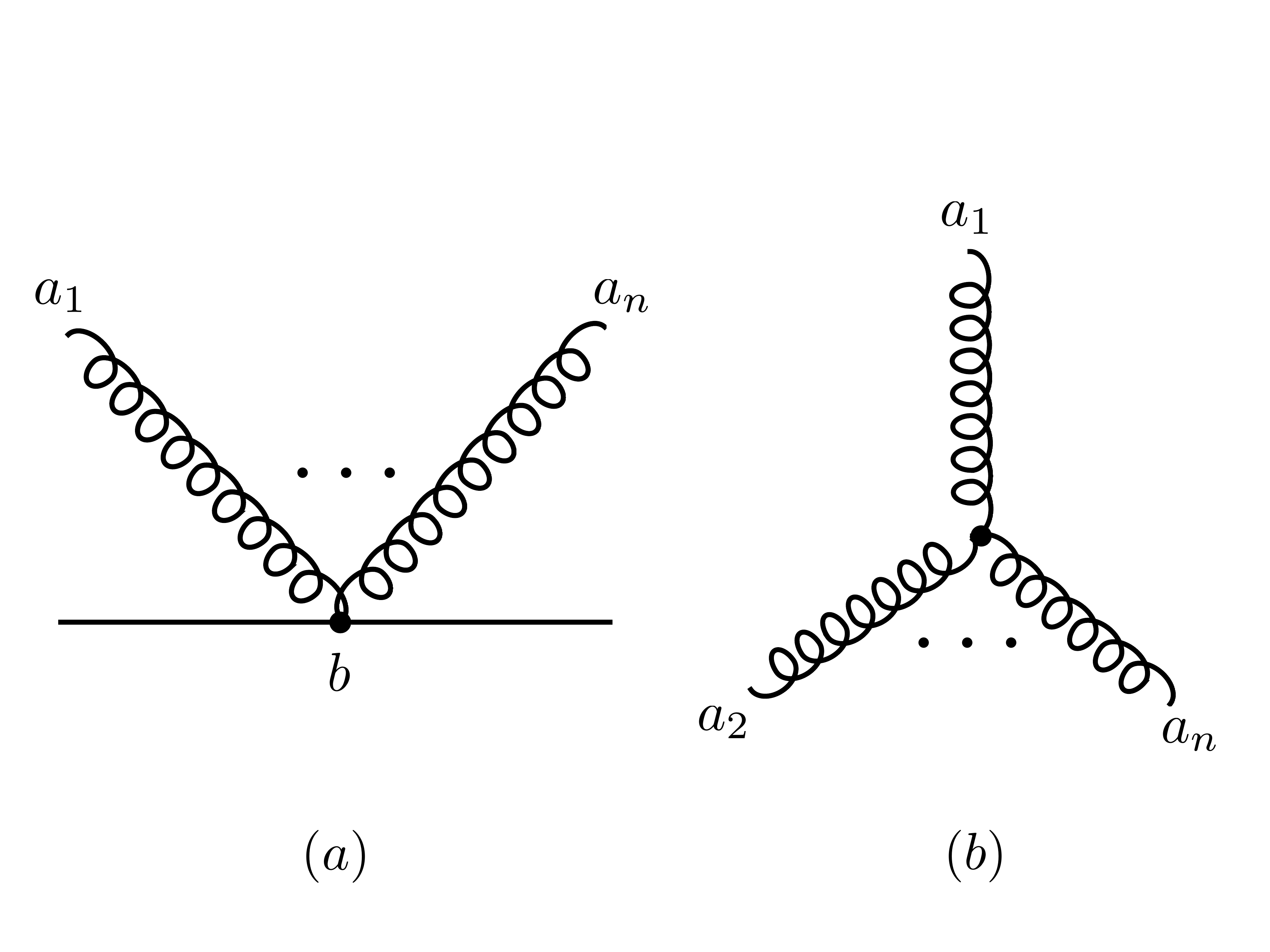}
    \caption{Interaction vertices. Diagram (a) shows the interaction vertex for $n$ gravitons, denoted by curly lines, coupling to a point particle, denoted by a straight line. Diagram (b) shows the self-interaction vertex of $n$ gravitons. The labels $a_1, a_2, \ldots$ and $b$ are CTP indices, which take values of $1$ and $2$.}
    \label{fig:Interactions}
\end{figure}

We now turn our attention to power counting the interaction terms in
(\ref{interactions1}). We consider first the diagram for the
interaction of $n$ gravitons with the effective particle worldline,
as shown in Fig.(\ref{fig:Interactions}a). The curly line denotes a
two-point function $D^{ab}_{\alpha \beta \gamma^\prime
\delta^\prime}$ of the metric perturbation (i.e. of a graviton). The
straight line denotes the effective point particle. We remark that
from the point of view of the gravitons, the particle acts as an
external source that couples to the metric perturbations. As such,
the straight line in Fig.(\ref{fig:Interactions}a) does {\it not}
represent the physical propagation of the compact object, but acts
as an external field, with the vertex denoting the particle-field
interactions. 

The power counting of $n$ gravitons interacting with
the effective particle is given by
\begin{eqnarray}
    {\rm Fig. ~(\ref{fig:Interactions}a)} = i S_{pp} ^{(n)} \sim \frac{ m}{ m_{pl}^n } \, d\tau \, h^n \sim \ve \left( \frac{ L }{ \ve } \right) ^{1-\frac{n}{2} }  .  \label{ppinterxns55}
\end{eqnarray}
The self-interaction vertices that result from the nonlinearity of the Einstein-Hilbert action are given in Fig.(\ref{fig:Interactions}b). The power counting for the self-interaction of $n$ gravitons gives
\begin{eqnarray}
    {\rm Fig. ~(\ref{fig:Interactions}b)} = i S^{(n)} \sim m_{pl}^2 d^4x \, \nabla^2 \frac{h^n}{ m_{pl}^n} \sim \left( \frac{ L }{ \ve } \right) ^{ 1- \frac{n}{2}  }  .
\end{eqnarray}
From Table \ref{table:scaling} we see that the power counting indicates that every type of interaction term involving any number of gravitons scales as $L^p$ where $p \le 1$.

\subsection{Feynman rules and calculating the effective action}
\label{ch3:Feynmanrules}

We now turn to calculating the effective action from (\ref{effaction0}) for $J^{\mu\nu}_a =0$. Standard quantum field theory arguments \cite{BirrellDavies, Peskin, Ryder} demonstrate that the effective action is given by
\begin{eqnarray}
    i \Gamma [ \langle z^\mu_a \rangle ] &=& i S_{pp}^{(0)} [ \langle z^\mu_a \rangle ] + \left(
        \begin{array}{c}
            {\rm sum ~of ~all ~1PI} \\
            {\rm  connected ~diagrams}
        \end{array}
        \right) . \nonumber \\
        &&
\end{eqnarray}
By ``connected diagrams" we mean those contiguous diagrams
constructed using the Feynman rules for the interaction terms in
(\ref{interactions1}). By ``1PI connected diagrams" we mean those connected diagrams that are one-particle-irreducible \cite{Peskin, Ryder}. However, we are only interested in those
connected diagrams that contribute at the classical level since the
quantum corrections due to graviton loops on the motion of an
astrophysical body are utterly negligible. Therefore, the effective
point particle worldline is assumed to be totally decohered
\footnote{There will always be non-zero worldline fluctuations, the
existence of which will be necessary for computing the semiclassical
equations of self-force. However, these fluctuations are so small
that only tree-level processes are relevant.} and we will simply
represent the expectation value of the worldline coordinates
operators $\langle \hat{z}_a^\alpha \rangle$ by their semiclassical values
$z_a^\alpha$.


A diagram with $\ell$ graviton loops scales as $L^{1-\ell}$, in units where $\hbar =1$. Therefore, classical processes correspond to those diagrams that scale linearly with $L$ (i.e., tree-level diagrams) and provide the dominant contributions to the effective action so that
\begin{eqnarray}
     i \Gamma [z_a] &=& i S_{pp}^{(0)} [ z_a] + \left(
        \begin{array}{c}
            {\rm sum ~of ~all ~} O(L) \\
            {\rm {\rm~connected ~diagrams} }
        \end{array}
        \right) \nonumber \\
        && + \left(
        \begin{array}{c}
            {\rm higher ~order ~1PI} \\
            {\rm graviton ~loop ~corrections}
        \end{array}
        \right)  .
\end{eqnarray}
In this manner we construct a systematic method for computing
the self-force equations order by order in $\ve$. 

The relationship between the connected diagrams, the interaction terms and the power counting is provided by the Feynman rules so that given a diagram at a given order in $\ve$ we can translate these into mathematical expressions. The Feynman rules are the following:
\begin{enumerate}
    \item A vertex represents the particle-field interaction $i V_{pp \, (n)} ^{ \alpha_1 \cdots \alpha_{2n} }$ or the graviton self-interaction $i V_{(n)}^{\alpha_1 \cdots \alpha_{2n} \rho_1 \rho_2 }$ as appropriate,
    \item Each endpoint and vertex is labeled by a CTP index, $a_i=\{1,2\}$, such that a factor of $1$ and $(-1)^{a_i+1}$ is associated with each endpoint and vertex, respectively.
    \item Include a factor of the graviton two-point function $D^{ab}_{\alpha \beta \gamma^\prime \delta^\prime} (x,x^\prime)$ connecting vertices labeled by CTP indices $a$ and $b$ at spacetime points $x$ and $x^\prime$,
    \item Sum over all CTP indices, integrate over proper time for each particle-field vertex, and integrate over all spacetime for each graviton self-interaction vertex,
    \item Multiply by the appropriate symmetry factor, $S$.
\end{enumerate}
The symmetry factor $S$ for a particular self-force diagram is found from the following expression
\begin{eqnarray}
	S = \frac{1}{ n_1! \, n_2 ! \cdots n_k ! }
\end{eqnarray}
where $k$ is the number of different types of vertices and $n_i$ is the multiplicity of a particular vertex appearing in the diagram such that $\sum_{i=1}^k n_i$ equals the total number of vertices. 
We will show how these rules are implemented as we continue.

\begin{figure}
    \center
    \includegraphics[width=4.5cm]{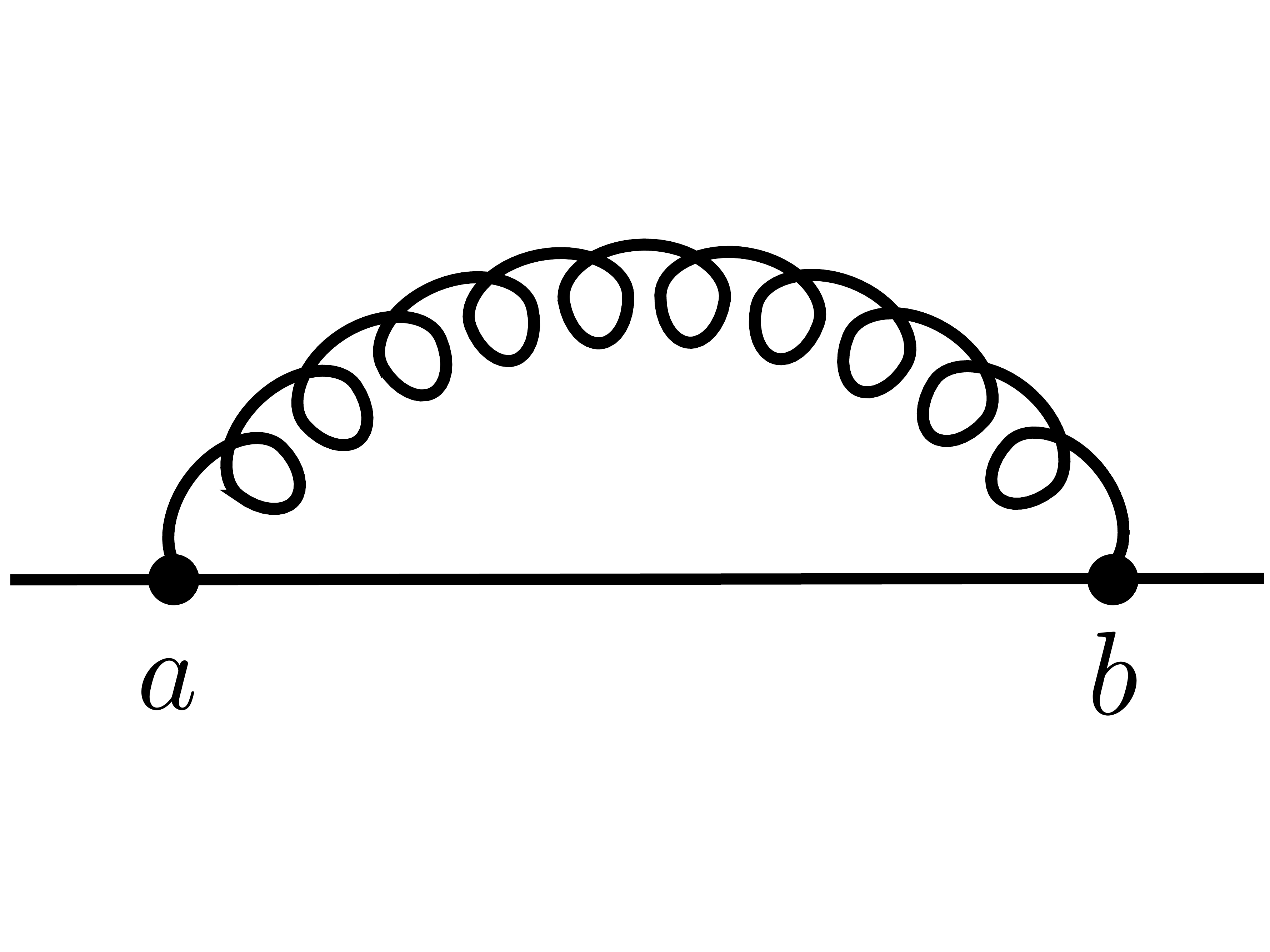}
    \caption{The diagram contributing to the first-order self-force described by the MST-QW equation.}
    \label{fig:firstordersf}
\end{figure}

To derive the MST-QW self-force equation we only need those diagrams
that contribute at $O( \ve L)$. From the Feynman rules for the
interactions in Fig.(\ref{fig:Interactions}) it follows that there is
only one such diagram at this order, which is shown in
Fig.(\ref{fig:firstordersf}). Therefore, the effective action to
first order in $\ve$ is
\begin{eqnarray}
    i \Gamma [ z_a ] = -i m \int d\tau_a + 
    {\rm Fig.(\ref{fig:firstordersf})} + O(\ve^2 L)
\end{eqnarray}
where
\begin{eqnarray}
    {\rm Fig.(\ref{fig:firstordersf})} &=& \left( \frac{1}{2!} \right) \left( \frac{i m}{2m_{pl}} \right)^2  \sum_{a,b=1}^2 \int d\tau_a \int d\tau_b ^\prime \, \nonumber \\
    && \times \, \dot{z}_a^\alpha \dot{z}_a^\beta  \, D^{ab} _{\alpha \beta \gamma^\prime \delta^\prime} \big( z_a^\alpha, z_b^{\alpha^\prime} \big)  \dot{z}_b^{\gamma^\prime} \dot{z}_b^{\delta^\prime} .   \label{firstordersf0}
\end{eqnarray}
The symmetry factor here is $1/2!$ since there are $n_1=2$ occurrences of the vertex $V_{pp}^{(1)}$ in the diagram.

Because the quantum fluctuations of the particle worldline are so strongly suppressed, it is reasonable to isolate the dominant contribution to the effective action and expand about $z_1 ^\alpha = z_2 ^\alpha$. Defining the difference and semi-sum (or averaged) coordinates
\begin{eqnarray}
    z_- ^\alpha &=& z_1 ^\alpha - z_2 ^\alpha \\
    z_+ ^\alpha &=& \frac{ z_1 ^\alpha + z_2 ^\alpha }{ 2 }
\end{eqnarray}
and implementing the identities in Appendix \ref{app:twoptfns} we find that such an expansion gives
\begin{eqnarray}
    i \Gamma [ z_a] &=& -i m \int d\tau \, z_-^\mu g_{\mu\alpha} a^\alpha_+  + \frac{i m^2}{2 m_{pl} ^2 } \int \! d\tau \! \int \! d\tau^\prime \, z_- ^\mu (\tau)  \nonumber \\
    && \times w_\mu ^{~\alpha \beta \nu} [z_+^\alpha] \nabla_\nu D^{ret}_{\alpha \beta \gamma^\prime \delta^\prime} (z^\alpha_+, z^{\alpha^\prime} _+ ) \dz_+^{\gamma^\prime} \dz_+^{\delta^\prime} + O(z_-^2) \nonumber \\
    &&    \label{effaction2}
\end{eqnarray}
where the $4$-acceleration is
\begin{eqnarray}
    a^\mu _+ (\tau) = \frac{D \dz^\mu _+ }{ d\tau}  ,
\end{eqnarray}
$\tau$ is the proper time associated with the worldline described by the semi-sum coordinates $z^\alpha_+$ so that
\begin{eqnarray}
    g_{\alpha \beta} (z_+ ) \dz ^\alpha _+ \dz^\beta_+ = -1
\end{eqnarray}
and the tensor $w^{\mu\alpha \beta \nu} [z]$ is defined by
\begin{eqnarray}
    w^{\mu\alpha \beta \nu} &=& \frac{1}{2} \, u^\alpha u^\beta w^{\mu \nu} - w^{\mu ( \alpha} u^{\beta )} u^\nu  \label{mpvecop0}  \\
    w^{\mu\nu} &=& g^{\mu\nu} + u^\mu u^\nu . \label{mpvecop1}
\end{eqnarray}

We remark that the CTP generating functional and the effective action provide causal dynamics for the effective particle's motion
since the retarded propagator is the only two-point function that
appears in (\ref{effaction2}).
Interestingly, however, the difference in using the in-out versus the
in-in formalisms is not overtly manifest at first order in $\ve$. We
will see an explicit example of these different approaches when we
calculate the second order self-force equations in the next paper.

We observe that the retarded propagator in (\ref{effaction2}) is
divergent when $\tau^\prime = \tau$. In order to have a finite and
well-behaved force on the compact object from the metric
perturbations we will need to regularize this divergence and possibly
renormalize the appropriate couplings of the theory.

\subsection{Regularization of the leading order self-force}
\label{ch3:reg}

The CS-EFT approach is founded in the theory of quantum fields in
curved spacetime \cite{BirrellDavies, Wald}. The renormalization of
divergences in this context has received much attention over the
decades and a considerable body of techniques has been developed to
remove these divergences in a systematic and self-consistent manner.
We therefore find it natural to renormalize the divergence in
(\ref{effaction2}) using these methods even if they are somewhat
unfamiliar in classical gravitational problems.

Of these approaches the method of dimensional regularization
\cite{tHooftVeltman:NuclPhysB44} is particularly useful. This
regularization scheme preserves the general coordinate and gauge
symmetries of the theory but is also a natural choice to use within
an effective field theory framework \cite{Burgess:EFT, Manohar:EFT,
Rothstein:EFT1}. The reason for the latter can be
seen from the problems that can develop when a simple cut-off
regularization is used for the divergent integrals appearing, for
example, in the Fermi four-point interaction theory of weak
interactions. We refer the reader to \cite{Manohar:EFT} for the
particular details of this theory. In this effective field theory the
mass of the W-boson $M_W$ is very large compared to the other masses
(e.g., quarks) and typical momenta in the problem so that the action
describing the low-energy theory is an expansion in powers of
$1/M_W$. As a result, using a momentum cut-off $\Lambda$ one finds
divergent diagrams at each order that scale like
\begin{eqnarray}
    \sim \left( \frac{ \Lambda }{ M_W } \right)^p
\end{eqnarray}
where $p$ is an integer. Since $\Lambda$ represents the scale at
which high energy physics becomes relevant it is natural to choose
$\Lambda \sim M_W$.
Therefore, all of the (power)  divergences at each order contribute
at $O(1)$ and the perturbative expansion in the EFT breaks down,
unless when they are resummed in a particular
manner \cite{Manohar:EFT}. This feature does not occur with dimensional regularization
since the dimensional parameter $\mu$ never shows up as a power
$\mu^p$ but appears only in logarithms. This is true of any
so-called mass-independent renormalization scheme \cite{Manohar:EFT}.

The smearing prescription developed and implemented in
\cite{GalleyHu:PRD72, GalleyHuLin:PRD74} to regularize the divergent part of the
retarded propagator should not be used within our CS-EFT
because it is a mass-dependent regularization scheme. This is easily
seen by looking at the shift in the mass of the electric point charge
(See Eq.(3.17) and the discussion following in \cite{GalleyHuLin:PRD74}) in which the divergence is
\begin{eqnarray}
    \frac{ \delta m }{ m } = N e^2 \frac{\Lambda}{m}
\end{eqnarray}
where $N$ is an $O(1)$ number and $\Lambda$ is a mass scale for smearing the divergence on the particle worldline. 
Matching onto the full (microscopic) theory implies that $1 / \Lambda$ is of order the ``classical radius" of the charge $r_o \sim e^2 / m$, which defines the length scale at which the vacuum polarization induced by the charge's presence becomes relevant. (Pair creation becomes important at this scale and marks roughly the length scale important for quantum electrodynamical processes \cite{LandauLifshitz}.) Then $\delta m / m$ is an $O(1)$ contribution thereby causing the perturbative expansion to break down since the high-energy physics no longer provides a small correction to low-energy processes. Therefore, the smearing regularization of \cite{GalleyHu:PRD72, GalleyHuLin:PRD74} is unsuitable for using within the CS-EFT framework.

Instead, we use the dimensional regularization  scheme below because
of its practical ease and because it allows for the effective field
theory to be renormalized in a manner consistent with the associated
perturbation series in $\ve$. Because we are applying a quantum field
theoretical renormalization scheme to a classical gravitational
problem we provide below a somewhat pedagogical discussion of the
steps in the regularization and renormalization of the effective
action.

The propagator is divergent in the limit when $x^\prime \to x$ and can be written as the sum of a regular and a divergent contribution \cite{BirrellDavies},
\begin{eqnarray}
    D = D^{ren} + D^{div}
\end{eqnarray}
(we are temporarily dropping the spacetime indices as well as the $ret$
label). The renormalized propagator is defined by the finite
remainder
\begin{eqnarray}
    D^{ren} \equiv D - D^{div}  = Pf ( D )  \label{proppseudofn0}
\end{eqnarray}
where $Pf$ stands for the pseudofunction of the quantity in
parentheses and is well-behaved as a (regular) distribution when
$x^\prime = x$. Generically, quantum two-point functions and
propagators are regarded as distributions and therefore only make
sense when integrated against a test function. Let us therefore
define $j(\lambda)$ to be such a testing function so that we can form
the (divergent) integral
\begin{eqnarray}
    {\cal I} \equiv \int_{-\infty} ^\infty d\lambda^\prime \, D ( z^\alpha, z^{\alpha^\prime} ) j (\lambda^\prime)  \label{divergentint0}
\end{eqnarray}
where we evaluate the propagator on the particle worldline $z^\alpha(\lambda)$. We refer the reader to Appendix \ref{app:distro} for our notations and definitions regarding distribution theory as well as to the excellent texts of \cite{Schwartz, Zemanian}.

The divergent integral in (\ref{divergentint0}) can be written as
\begin{eqnarray}
    {\cal I} = \lim _{\epsilon \to 0} \Big[ I (\epsilon ) + H(\epsilon) \Big]  \label{divergentint1}
\end{eqnarray}
where $I(\epsilon)$ is the divergent part of the integral and $H(\epsilon)$ is the finite part. These are related to the renormalized and divergent propagators through
\begin{eqnarray}
    \lim _{\epsilon \to 0} I(\epsilon) &=& \int _{-\infty} ^\infty d\lambda^\prime \, D^{div} (z ^\alpha, z^{\alpha^\prime} ) j(\lambda^\prime) \\
    H(0) &=& \int_{-\infty} ^\infty d\lambda^\prime \, D^{ren} (z^\alpha, z^{\alpha^\prime} ) j(\lambda^\prime)  \\
        &=& Fp \int_{-\infty} ^\infty d\lambda^\prime \, D (z^\alpha, z^{\alpha^\prime} ) j(\lambda^\prime)
\end{eqnarray}
where $Fp$ denotes the finite part of the divergent integral, in the sense of Hadamard \cite{Hadamard}, in (\ref{divergentint0}). These relations follow from the fact that the renormalized propagator in (\ref{proppseudofn0}) is a pseudo-function and, by definition, generates the finite part of (\ref{divergentint0}).

In general, as we  discuss in Appendix \ref{app:distro}, the divergent part can be written in terms of its power divergent terms as well as powers of logarithms \cite{Schwartz, Zemanian}
\begin{eqnarray}
    I(\epsilon) = \sum_{p=1}^N \frac{a_p}{\epsilon^p} + \sum_{p=1}^M b_p \ln^p \epsilon  \label{divprop0}
\end{eqnarray}
as $\epsilon \to 0$. Here $\epsilon$ is related to a cut-off momentum, $\epsilon \sim 1/\Lambda$, and is used for illustrative purposes. To identify the divergent part of the integral in (\ref{divergentint0}) we need a representation for $D^{div}$ that allows for a clear separation of the divergent parts from the finite terms.

We use a momentum space representation for the graviton propagator
that is initially introduced by Bunch and Parker for a scalar field
in an arbitrary curved spacetime in \cite{BunchParker:PRD20}. Keeping only those terms in $D^{div}$ that are divergent
amounts to expanding the propagator up to a specific order  ($n^{\rm th}$) in
derivatives of the background metric when using Riemann normal
coordinates. Throwing away all higher derivative terms in the
expansion, which are ultraviolet finite as can be verified by power
counting the momentum integrals, results in the divergent structure
shown in (\ref{divprop0}). We may therefore write
\begin{eqnarray}
    D^{div} = D^{BP}_{(n)}
\end{eqnarray}
where the superscript $BP$ stands for the Bunch-Parker momentum space
representation.

Returning to (\ref{divergentint1}), the finite part of the integral is defined via the pseudofunction of the propagator in (\ref{proppseudofn0}) so that
\begin{eqnarray}
    && {\hskip-.5in} Fp \int_{-\infty}^\infty d\lambda^\prime \, D(z^\alpha, z^{\alpha^\prime} ) j(\lambda^\prime) \nonumber \\
    &=& \lim_{\epsilon \to 0} \left( \int_{-\infty} ^{\lambda - \epsilon} + \int_{\lambda+\epsilon} ^\infty \right) d\lambda^\prime \, D (z^\alpha, z^{\alpha^\prime} ) j(\lambda^\prime)  \nonumber \\
    && - \int_{-\infty} ^\infty d\lambda^\prime \, D^{BP}_{(n)} (z^\alpha, z^{\alpha^\prime} ) j(\lambda^\prime) .
     \label{finpart0}
\end{eqnarray}
We may then write the worldline integral of the full propagator as
\begin{eqnarray}
    {\cal I} &=& Fp \int_{-\infty} ^\infty d\lambda^\prime \, D ( z^\alpha, z^{\alpha^\prime} ) j(\lambda^\prime) \nonumber \\
    && + \int_{-\infty} ^\infty d\lambda^\prime \, D^{BP}_{(n)} ( z^\alpha, z^{\alpha^\prime} ) j(\lambda^\prime)  .
\end{eqnarray}
Using these expressions, we find that the first order self-force in (\ref{effaction2}) is given by
\begin{eqnarray}
    {\rm Fig.(\ref{fig:firstordersf})} &\arreq& \frac{i m^2 }{ 2 m_{pl}^2 } \int_{-\infty}^\infty \!\!  d\tau \, z^\mu _- \, w_\mu ^{~\alpha \beta \nu} [z_+^\alpha] \nabla_\nu \nonumber \\
    && \Bigg\{ Fp \int_{-\infty} ^{\infty} \!\! d\tau^\prime \, D^{ret}_{\alpha \beta \gamma^\prime \delta^\prime} \big( z_+^\alpha, z_+^{\alpha^\prime} \big)  \dz_+^{\gamma^\prime} \dz_+^{\delta^\prime}  \nonumber \\
    && ~~~ + \int_{-\infty} ^{\infty} \!\!  d\tau^\prime \, D ^{BP} _{(n) \alpha \beta \gamma^\prime \delta^\prime} \big( z_+^\alpha, z_+^{\alpha^\prime} \big)  \dz_+^{\gamma^\prime} \dz_+^{\delta^\prime}  \Bigg\} \nonumber \\
    &&
\end{eqnarray}
where we parameterize the worldline by the particle's proper time.

Ignoring terms that are higher order in $m$, 
the divergent contribution that arises from the second term is
\begin{eqnarray}
     I_\mu ^{~\nu} (\tau) &=& w_{\mu}^{~\alpha \beta \nu} [z_+^\alpha] \int_{-\infty} ^{\infty} \!\!\! d\tau^\prime \, D ^{BP} _{(n) \alpha \beta \gamma^\prime \delta^\prime} \big( z_+^\alpha, z_+^{\alpha^\prime} \big)  \dz_+^{\gamma^\prime} \dz_+^{\delta^\prime}  . \nonumber \\
     && \label{divintegral0}
\end{eqnarray}
The particle worldline is a geodesic of the background spacetime at leading order so that the velocity at $\tau^\prime$ is related to that at proper time $\tau$ through
\begin{eqnarray}
    \dz_+^{\gamma^\prime} (\tau^\prime) = g^{\gamma^\prime} _{~\mu} \big( z_+(\tau^\prime), z_+(\tau) \big) \dz_+^\mu (\tau)
\end{eqnarray}
where $g^{\gamma^\prime} _{~\mu}$ is the bi-vector of parallel transport, which parallel transports a vector at $z_+(\tau)$ to another point $z_+(\tau^\prime)$ along the unique geodesic connecting these points, namely, the leading order worldline of the effective particle's motion. The divergent integral (\ref{divintegral0}) can then be written as
\begin{eqnarray}
    I_\mu ^{~\nu} (\tau) &\arreq& w_\mu ^{~\alpha \beta \nu} [z_+^\alpha] \dz_+^\rho  \dz_+^\sigma \int_{-\infty } ^\infty d\tau^\prime \, D^{BP} _{(n) \alpha \beta \gamma^\prime \delta^\prime} \big( z_+^\alpha, z_+^{\alpha^\prime} \big)  \nonumber \\
    && ~~~ \times g^{\gamma^\prime} _{~\rho} \big( z_+^{\alpha^\prime}, z_+^\alpha \big) g^{\delta^\prime} _{~\sigma} \big( z_+^{\alpha^\prime}, z_+^\alpha \big)  .
\end{eqnarray}
The integrand is now a rank-4 tensor at $z_+(\tau)$ and a scalar at $z_+(\tau^\prime)$, which we write as
\begin{eqnarray}
    D^{BP}_{(n) \alpha \beta \gamma \delta} \big( z_+^\alpha, z_+^{\alpha^\prime} \big) &\equiv& D^{BP} _{(n) \alpha \beta \gamma^\prime \delta^\prime} \big( z_+^\alpha, z_+^{\alpha^\prime} \big) \nonumber \\
    && \times g^{\gamma^\prime} _{~\gamma} \big( z_+^{\alpha^\prime}, z_+^\alpha \big) g^{\delta^\prime} _{~\delta} \big( z_+^{\alpha^\prime}, z_+^\alpha \big)   . \nonumber \\
    &&
\end{eqnarray}

In \cite{Galley:momentum} we derive the momentum space representation of the Feynman propagator for metric perturbations, which is found to be
\begin{eqnarray}
    && {\hskip-0.2in} D^{BP} _{(2) \hata \hatb \hatc \hatd} (y) =  -i \int _{ {\cal C}_{ret} } \!\! \frac{d^d k}{ (2\pi)^d } \, e^{ik\cdot y} \Bigg\{ \frac{P_{\hata \hatb \hatc \hatd} (\eta) }{k^2}  + \frac{ 2 R_{\hata (\hatc \hatd) \hatb } }{ k^4 }   \nonumber \\
    &&- \frac{2}{3} \, \frac{ k^\hats k^\hatt }{ k^6 } \Bigg[ \eta_{\hata \hatc} R_{\hatb \hats \hatd \hatt} + \eta_{\hata \hatd} R_{\hatb \hats \hatc \hatt}  + \eta_{\hatb \hatc} R_{\hata \hats \hatd \hatt} + \eta_{\hatb \hatd} R_{\hata \hats \hatc \hatt}  \nonumber \\
    && + \frac{ 4(d-3) }{ (d-2)^2 }  \Big( \eta_{\hata \hatb} R_{\hatc \hats \hatd \hatt} + \eta_{\hatc \hatd} R_{\hata \hats \hatb \hatt} \Big) \Bigg] \Bigg\}  \label{BP2}
\end{eqnarray}
where we have ignored those terms that fall off as $k^{-5}$ in the integrand since these give finite contributions in $d=4$ dimensions, $y$ denotes the Riemann normal coordinates (RNC) of $z^\alpha (\tau^\prime)$ with respect to the origin at $z^\alpha (\tau)$ and ${\cal C}_{ret}$ is the contour appropriate for the retarded propagator. The hatted indices represent tensor components evaluated in Riemann normal coordinates.

Using (\ref{BP2}) the divergent integral (\ref{divintegral0}) in RNC is
\begin{eqnarray}
    I^{\hatm \hatn} (\tau) &=& -i \int_{-\infty}^\infty \!\!\! d\tau^\prime \int _{\cC_{ret}} \!\! \frac{ d^d k }{ (2\pi)^d} \, e^{ik \cdot y} \Bigg\{ \frac{1}{2} \, \frac{d-3}{d-2} \, \frac{ w^{\hatm \hatn} }{ k^2 } \nonumber \\
    && - \frac{8}{3} \, \frac{d^2 - 6d +10}{ (d-2)^2 } \, \frac{ k^\hats k^\hatt }{ k^6 } \, R_{\hata \hats \hatb \hatt} \eta^{\hatm ( \hatn} \dz^{\hata ) } \dz^\hatb \Bigg\} \nonumber \\
    && \label{divintegral01}
\end{eqnarray}
where we have used $\dz^\hatb \dz^\hatc \dz^\hatd R_{\hata\hatc\hatb\hatd}=0$. 
We provide the computational details to regularize $I^{\hatm \hatn}$ in Appendix \ref{app:regularize} using dimensional regularization and demonstrate that it vanishes.

The remaining finite part of Fig.(\ref{fig:firstordersf}) is therefore
\begin{eqnarray}
    {\rm Fig.(\ref{fig:firstordersf})} &\arreq& \frac{ i m^2 }{ 2 m_{pl}^2 } \int d\tau \, z_-^\mu w_\mu ^{~\alpha \beta \nu} [z_+] \nabla_\nu   \nonumber \\
    && Fp \int_{-\infty} ^{\infty} d\tau^\prime \, D^{ret}_{\alpha \beta \gamma^\prime \delta^\prime} \big( z_+^\alpha, z_+^{\alpha^\prime} \big)  \dz_+^{\gamma^\prime} \dz_+^{\delta^\prime}  .  ~~ \label{finintegral0}
\end{eqnarray}
Notice that we do not need to renormalize any parameters of the
theory at this order since dimensional regularization sets the power
divergence to zero.

Having regularized the leading order contribution to the effective
action in Fig.(\ref{fig:firstordersf}) we now compute the equations
of motion from (\ref{finintegral0}). The variational principle
\begin{eqnarray}
    \frac{ \delta \Gamma [z_a] }{ \delta z_- ^\mu (\tau) } \Bigg|_{z_- = 0 } = 0
\end{eqnarray}
yields
\begin{eqnarray}
    m a^\mu (\tau) &=& \frac{ m^2 }{2 m_{pl}^2 } w^{\mu \alpha \beta \nu} [z^\alpha] \nabla_\nu   \nonumber \\
    && Fp \int _{-\infty} ^\infty d\tau^\prime \, D^{ret} _{\alpha \beta \gamma^\prime \delta^\prime} (z^\alpha, z^{\alpha^\prime} ) \, \dz^{\gamma^\prime} \dz^{\delta^\prime}  . ~~~~~~
\end{eqnarray}
Using the definition of Hadamard's finite part of the integral in (\ref{finpart0}) and the fact that the retarded propagator is zero for $\tau^\prime > \tau$ we see that the finite part is given by
\begin{eqnarray}
    && {\hskip-0.4in} Fp \int_{-\infty} ^\infty d\tau^\prime \, D^{ret} _{\alpha \beta \gamma^\prime \delta^\prime} (z^\alpha, z^{\alpha^\prime} ) \, \dz^{\gamma^\prime} \dz^{\delta^\prime}   \nonumber \\
    && = \lim _{\epsilon \to 0} \int_{-\infty} ^{\tau- \epsilon } d\tau^\prime \, D^{ret} _{\alpha \beta \gamma^\prime \delta^\prime} (z^\alpha, z^{\alpha^\prime} ) \, \dz^{\gamma^\prime} \dz^{\delta^\prime}  .
\end{eqnarray}
Inserting
this into the equations of motion \footnote{We also pass the
covariant derivative $\nabla_\nu$ through the integral. In
\cite{GalleyHuLin:PRD74} we show that this can be done without
introducing local and conservative forcing terms, at least in the gravitational
context.} gives the equation for the self-force on the compact object moving in a vacuum background spacetime
\begin{eqnarray}
    a^\mu (\tau) &=& \frac{ m }{2 m_{pl}^2 } w^{\mu \alpha \beta \nu} [z^\alpha]   \nonumber \\
    && \times \lim_{\epsilon \to 0} \int _{-\infty} ^{\tau-\epsilon} {\hskip-0.15in} d\tau^\prime \, \nabla_\nu D^{ret} _{\alpha \beta \gamma^\prime \delta^\prime} (z^\alpha, z^{\alpha^\prime} ) \, \dz^{\gamma^\prime} \dz^{\delta^\prime}  ,   \nonumber \\
    && \label{firstsf0}
\end{eqnarray}
which was originally derived by Mino, Sasaki and Tanaka \cite{MinoSasakiTanaka:PRD55} and by Quinn and Wald \cite{QuinnWald:PRD56}.

\section{Effacement Principle for EMRIs}
\label{sec:finsize}

While it is intuitive to expect finite size corrections to be
negligibly small at the linear order self-force one may be concerned
with such corrections at higher orders. Specifically, at what order
in $\ve$ is the motion of the compact object affected by induced
tidal deformations? In this section we answer this question using
coordinate invariant arguments and demonstrate that such finite size
effects from a small Schwarzschild black hole or neutron star moving
in a background curved spacetime unambiguously enter the self-force
at $O(\ve^5)$ and as deviations from a point particle motion at
$O(\ve^4)$. We also discuss the tidal deformations of a white dwarf star, which are more sensitive to the curvature of the background geometry. 

\subsection{Non-spinning black holes and neutron stars}

To begin we write down the effective point particle action that includes all possible self-interaction terms consistent with general coordinate invariance and worldline reparameterization invariance,
\begin{eqnarray}
    S_{pp} [z]  &=& - m \int d\tau + c_E \int d\tau \, \cE_{\mu\nu} \cE^{\mu\nu}   \nonumber \\
    && + c_B \int d\tau \, \cB_{\mu\nu} \cB^{\mu\nu} + \cdots   \label{effpp2}
\end{eqnarray}
where we have already used a field-redefinition to remove those terms
involving the Ricci curvature tensor. The terms containing the square of
the Riemann curvature (and higher powers) represent the influence of
the finite size of the body as it moves through space. This is seen
by noting that the equations of motion no longer have vanishing
acceleration so that the effective point particle
does not move along a geodesic of the spacetime,
\begin{eqnarray}
	m a^\mu (\tau) = 2 c_E \cE_{\alpha \beta} \cE^{\alpha \beta ; \mu} + \cdots  .
\end{eqnarray}
Such deviation from geodesic motion is
typical of tidally-distorted bodies.

The coefficients $c_{E,B}$ are parameters that depend upon the structure of the extended body. We must therefore match the effective point particle theory onto the full theory in order to encode this ``microscopic" or ``high-energy" structure onto the long wavelength effective theory. The matching procedure involves calculating (coordinate invariant) observables in both the effective theory and the full theory \footnote{Strictly speaking, one can use any quantity for the matching procedure but it is simpler to draw coordinate invariant conclusions by matching with a coordinate invariant quantity, such as a scattering amplitude, a cross-section, etc.}. By expanding the observable of the full theory in the long wavelength limit, where the effective theory is applicable, we can simply read off the values of $c_{E,B}$ as well as any other coefficients in (\ref{effpp2}). We outline the steps for a matching calculation and perform
an order of magnitude estimation 
of $c_{E,B}$ for a spherically symmetric compact object. 

\begin{figure}
    \center
    \includegraphics[width=4.5cm]{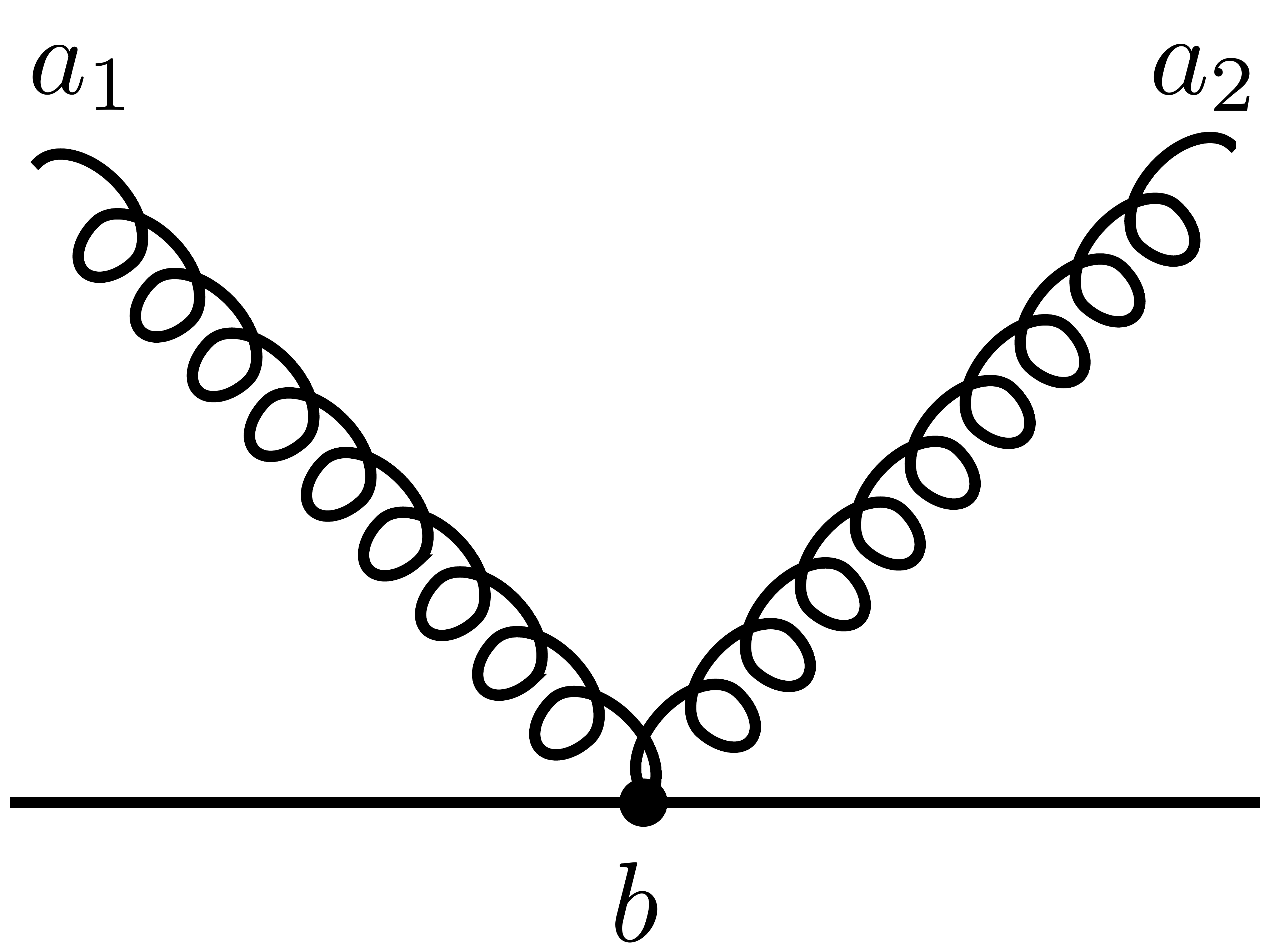}
    \caption{Graviton scattering off the background of a static and spherically symmetric extended body (e.g., a Schwarzschild black hole, a non-spinning neutron star).}
    \label{fig:compton}
\end{figure}

Below, we power count the scattering cross-section for graviton Compton scattering shown in Fig.(\ref{fig:compton}), which represents the scattering of gravitational waves in the spacetime of the isolated compact object.

We first compute the cross-section in the effective point particle theory. The scattering amplitude is computed by expanding the terms in (\ref{effpp2}) proportional to $c_{E}$ and $c_B$ to second order in the metric perturbations,
\begin{eqnarray}
    S_{pp}[ z ]  \!\! &=& \!\! \cdots + c_E \! \int d\tau \, \Big( \cE^{(0)}_{\mu\nu} \cE^{(0) \mu\nu} + 2 \cE_{\mu\nu} ^{(1)} \cE ^{(0) \mu\nu}   \nonumber \\
    && ~~~ + \cE_{\mu\nu} ^{(1)} \cE^{(1) \mu\nu} + 2 \cE_{\mu\nu} ^{(2)} \cE^{(0) \mu\nu} + \cdots  \Big) \nonumber \\
    && + c_B \! \int d\tau \, \Big( \cB^{(0)}_{\mu\nu} \cB^{(0) \mu\nu} + 2 \cB_{\mu\nu} ^{(1)} \cB^{(0) \mu\nu}   \nonumber \\
    &&  ~~~ + \cB_{\mu\nu} ^{(1)} \cB^{(1) \mu\nu} + 2 \cB_{\mu\nu} ^{(2)} \cB^{(0) \mu\nu} + \cdots \Big)   \nonumber \\
    &&  + \cdots   ,
\end{eqnarray}
where the superscript denotes the number of metric perturbations appearing in that function so that $\cB^{(2)}_{\mu\nu}$ is proportional to $h_{\alpha \beta} h_{\gamma \delta}$, for example. From the power counting rules developed in Section \ref{ch3:powercounting} we find that the scattering amplitude associated with Fig.(\ref{fig:compton}) scales as
\begin{eqnarray}
    i {\cal A} \sim \cdots \& ~ \frac{ c_{E,B} }{ m_{pl}^2 } \left( \frac{1}{\cR^2} \right)^2 ~ \& \cdots
\end{eqnarray}
where the $1/\cR^2$ comes from the two spacetime derivatives in the
Riemann tensor and the $\&$ denotes ``and a term that scales as".
While the cross-section includes contributions from other terms in
the effective particle action it will contain one term proportional
to $c_{E,B}^2$,
\begin{eqnarray}
    \sigma_{pp} \sim | i {\cal A} | ^2 \sim \cdots \& ~ \frac{ c^2 _{E,B} }{ m_{pl}^4 } \, \frac{ 1}{ \cR^8 } ~ \& \cdots
\end{eqnarray}
where the $pp$ subscript indicates that this is the cross-section computed in the effective point particle theory.

We turn now to the scattering cross-section in the full theory. A cross-section represents an effective scattering area  and the only scale present in the full theory of the isolated compact object is set by its size $r_m \sim m/ m_{pl}^2$. It follows that
\begin{eqnarray}
    \sigma_{full} = r_m^2 f \left( \frac{r_m}{\cR} \right)
\end{eqnarray}
where $f$ is a dimensionless function. In the long wavelength limit where $r_m / \cR \ll 1$ the cross-section will contain a term proportional to $\cR^{-8}$,
\begin{eqnarray}
    \sigma_{full} \sim \cdots \& ~ r_m^2 \left( \frac{ r_m }{ \cR} \right)^8  \& \cdots   .
\end{eqnarray}
Since quantities computed in the effective theory ought to match those computed in the long wavelength limit of the full theory we conclude that
\begin{eqnarray}
    c_{E,B} \sim m_{pl}^2 r_m^5 \sim \frac{ m^5 }{ m_{pl}^8}   \label{cEB}
\end{eqnarray}
upon identifying the $\cR^{-8}$ terms in both $\sigma_{pp}$ and $\sigma_{full}$.

\begin{figure}
    \center
    \includegraphics[width=7.5cm]{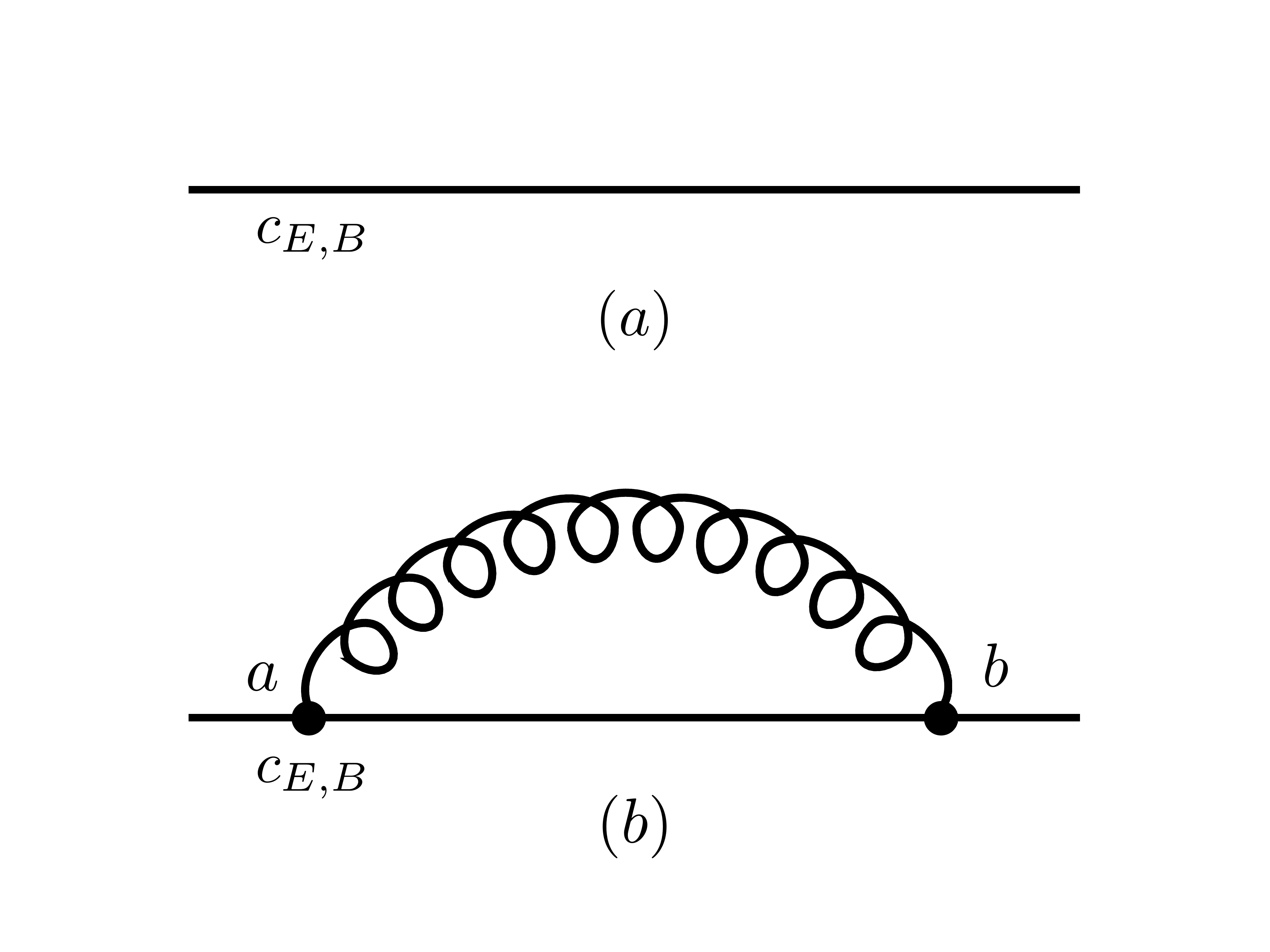}
    \caption{Lowest order contributions to (a) deviation from (minimal) point particle motion due to the tidal deformations of the compact object and (b) the self-force from the interaction of gravitational radiation with these deformations.}
    \label{fig:finsize}
\end{figure}

Using (\ref{cEB}) we can estimate the order in $\ve$ that the non-minimal terms appearing in the effective point particle action (\ref{effpp2}) will affect the motion of the compact object. The first diagram that the finite size terms will contribute is proportional to $c_{E,B}$ and is shown in Fig.(\ref{fig:finsize}a). This describes the deviation from the pure point particle motion experienced by the compact object due to the inclusion of the non-minimal couplings to the background spacetime. This diagram scales with $\ve$ as
\begin{eqnarray}
    {\rm Fig.(\ref{fig:finsize}a)} \sim c_{E,B} d\tau \left( \frac{1}{\cR^2} \right)^2 \sim \ve^4 L
\end{eqnarray}
and enters at fourth order.

This diagram does not couple to metric perturbations; it persists in the absence of gravitational radiation. As a result, while Fig.(\ref{fig:finsize}a) will affect the motion of the particle it is not a correction to the self-force. To find the order at which the tidal deformations affect the self-force we power count the diagram in Fig.(\ref{fig:finsize}b) to find that
\begin{eqnarray}
    {\rm Fig.(\ref{fig:finsize}b)} \sim c_{E,B} d\tau \left( \frac{1}{\cR^2} \right)^2 \frac{h}{m_{pl} } \sqrt{\ve L} \sim \ve^5 L   .
\end{eqnarray}
Finite size effects therefore enter the self-force at fifth order in $\ve$.

\subsection{Non-spinning white dwarf stars}

In (\ref{cEB}) we assume that the size of the compact object is of
the same order as its mass, $r_m \sim m/m_{pl}^2$. While this holds
true for black holes and neutron stars it does not for white dwarf
(WD) stars. WDs are thousands of times larger than their
Schwarzschild radius and subsequently experience stronger tidal
effects than a BH or NS with the same mass. In fact, the tides may be so severe that the WD is tidally disrupted at some point along its orbit about the SMBH. As a result, we expect that finite size effects may be (numerically) enhanced and alter the WD's motion, possibly at a lower order in $\ve$.

To see how this can arise we define the ratio of the compact object's radius to its mass by $f_{co}$ so that
\begin{eqnarray}
    r_m = f_{co} G m = \frac{ f_{co} }{ 32 \pi } \, \frac{ m }{ m_{pl}^2 }   \label{radiusCO1}
\end{eqnarray}
and the bookkeeping parameter $\ve$ becomes
\begin{eqnarray}
    \ve \equiv \frac{r_m}{ \cR} \sim f_{co} \, \frac{ m }{ m_{pl}^2 \cR }   \label{muSchw0}
\end{eqnarray}
with $m/m_{pl} \sim \sqrt{ \ve L/ f_{co} }$. If the compact object is a black hole then $f_{bh} = 2$. For a neutron star having a mass of $1.4 M_\odot$ and a radius between $10$ and $16$ km it follows that $f_{ns}$ varies between $4.8$ and $7.7$, respectively.

Using the new expressions for $r_m$ and $\ve$ it is not difficult to show that
\begin{eqnarray}
    c_{E,B} \sim f_{co} ^5 \frac{ m^5 }{ m_{pl}^8} \sim m_{pl}^2 r_m^5  \label{newscaling0}
\end{eqnarray}
and therefore the leading order finite size effects from induced tides scales as
\begin{eqnarray}
    {\rm Fig.(\ref{fig:finsize}a)} \sim f_{co} \ve^4 L .  \label{newscaling1}
\end{eqnarray}
Since $\ve \propto f_{co}$ it follows that Fig.(\ref{fig:finsize}a) scales as $f_{co}^5$. We remark that this observation agrees with that in \cite{FlanaganHinderer:0709,Thorne:PRD58} where the authors observe that the induced quadrupole moment of a neutron star scales as $r_{ns}^5$.

The parameter $\ve$ generally depends on some distance scale, $r$,  set by the orbit of the compact object. As a result, $\ve$ takes on different values at different orbital scales. Similarly, in the post-Newtonian expansion the small parameter (the relative velocity of the compact objects $v$) depends upon the scale of the orbital separation $r$ through the virial theorem, $v(r) \sim ( G m / r)^{1/2}$. 

To determine how the leading order finite size effects from Fig.(\ref{fig:finsize}a) depend on the orbital scale $r$ we consider the SMBH background to be a Schwarzschild black hole. It then follows from (\ref{muSchw0}) and (\ref{curvinvariant0}) that
\begin{eqnarray}
    \ve = \frac{r_m}{ \cR} = 2^{3/4} 3^{1/4} f_{co} \frac{ m}{ M} \left( \frac{r}{M} \right)^{-3/2}  .
\end{eqnarray}
As the orbital scale decreases we observe that $\ve$ increases.

For a WD, $\ve$ can be significantly larger than for a BH or NS with the same mass because $f_{wd}$ is typically much larger than both $f_{bh}$ and $f_{ns}$. In fact, $f_{wd}$ can be so large that Fig.(\ref{fig:finsize}a) may be {\it numerically enhanced} and important at orders in $\ve$ {\it lower} than the naive fourth order estimation given in the previous section. Such enhancement is consistent with our physical intuition that a WD experiences stronger tidal deformations than a BH or NS with the same mass.

To demonstrate this enhancement consider a simple example. Assume that the radius of the WD equals the Schwarzschild radius of the non-spinning SMBH it orbits. Choosing the radii to be $6000$ km each it follows that the masses of the WD and SMBH are $m \approx 0.6 M_\odot$ \cite{ShapiroTeukolsky} and $M=4000 M_\odot$, respectively. For these values, (\ref{radiusCO1}) indicates that
\begin{eqnarray}
    f_{wd} \approx 6750 ,
\end{eqnarray}
which is much larger than both $f_{bh}$ and $f_{ns}$, as expected.

\begin{figure}
    \center
    \includegraphics[width=8.5cm]{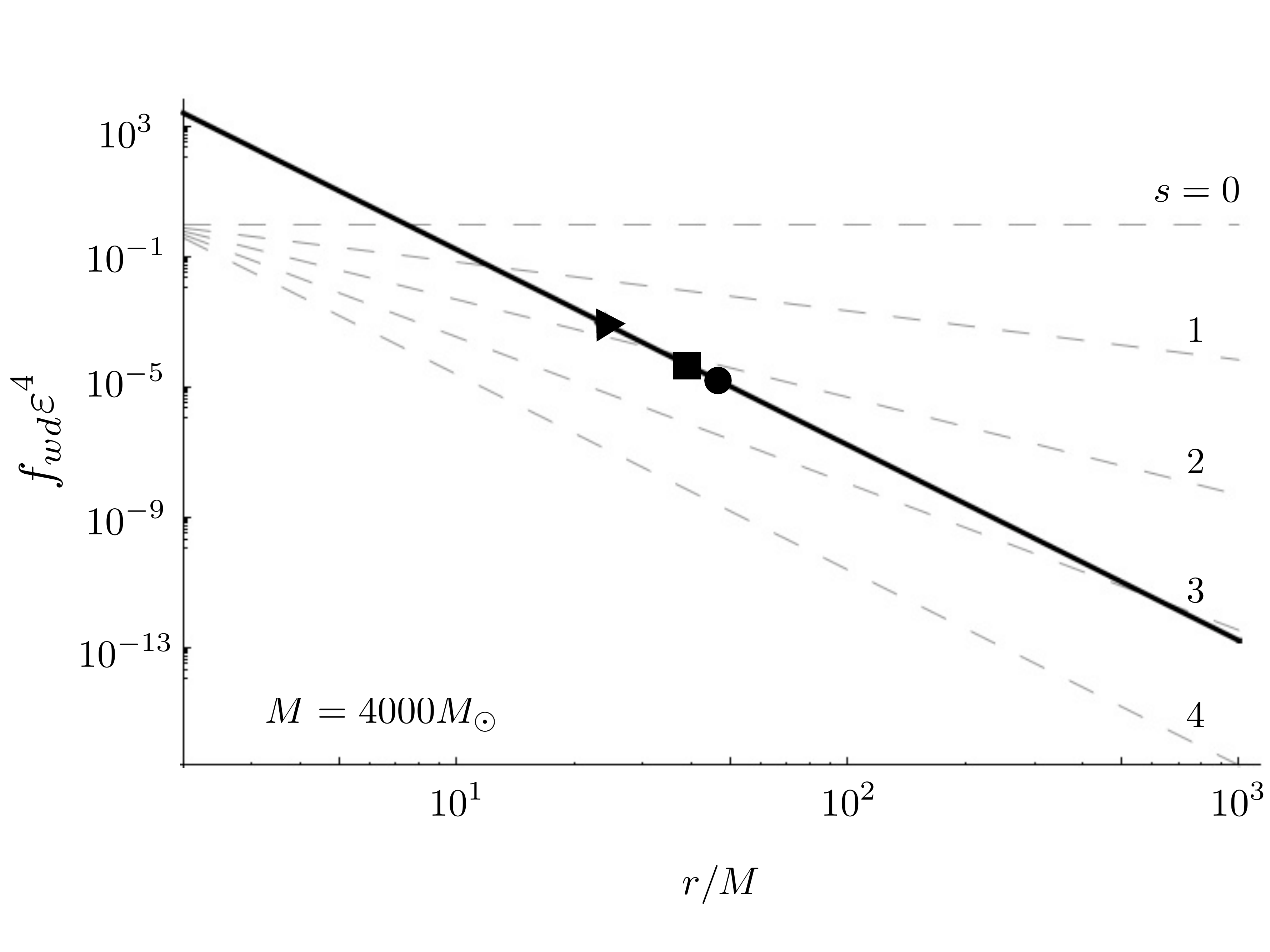}
    \caption{The effects from the finite size of a white dwarf star can be enhanced as it orbits in closer to the SMBH. The white dwarf seems to undergo some form of tidal disruption by either tidal disintegration (triangle and circle) or Roche lobe overflow (square). 
    In either case, tidal disruption may be numerically equivalent to a second order process.}
    \label{fig:scalecrossing}
\end{figure}

For this scenario, how do the leading order finite size effects of (\ref{newscaling1}) change with the orbital scale, $r$, of the WD? Fig.(\ref{fig:scalecrossing}a) shows a log-log plot of $f_{wd} \ve^4$ as a function of the radial distance $r/M$ from the SMBH. The dark line is $f_{wd} \ve^4$ while the dashed gray lines are plots of $\ve^s$ for $s=0, \ldots, 4$. As the orbital scale decreases, the dark line crosses several dashed gray lines indicating that Fig.(\ref{fig:finsize}a) is numerically enhanced from the naive $\ve^4$ estimate for BH's and NS's to order $\ve^s$ because of the stronger WD tidal deformations. It is the scaling of the non-minimal couplings $c_{E,B}$ in (\ref{newscaling0}) with the fifth power of $f_{wd}$ that is responsible for this enhancement of the leading order finite size effects.

However, this enhancement can not proceed indefinitely. If the tidal deformations become strong enough to transfer material from the WD to the SMBH (via Roche lobe oveflow) or to disintegrate the WD at its Roche limit then the WD is tidally disrupted and can no longer be described effectively as a point particle. This indicates that our particular construction of a point mass effective field theory will no longer satisfactorily describe the binary.

Using Newtonian estimates we find the Roche limit for the rigid (fluid) WD to be near $r \approx 24 M$ ($r \approx 47 M$). These scales are represented by a triangle (circle) in Fig.(\ref{fig:scalecrossing}). The WD's Roche lobe begins to overflow and transfer mass that accretes onto the SMBH when $r \approx 40M$ \cite{Eggleton:AstrophJ} and is denoted with a square in Fig.(\ref{fig:scalecrossing}). In all three cases the WD is tidally disrupted, a process that occurs near the $s=2$ line.

All of this suggests that finite size effects from induced tidal deformations can be enhanced for a WD until the star undergoes tidal disruption at $O(\ve^2)$. If one is interested in calculating the second order self-force on  a WD these tidal effects may have to be taken into account at some point during the binary's evolution to accurately determine the gravitational waveforms.


If the SMBH mass is increased to $10^5 M_\odot$
we estimate that tidal disruption occurs much closer to the SMBH's horizon. Describing the WD as an effective point particle is therefore valid over much of its orbital
evolution. Increasing the mass of the SMBH further we find that WD
tidal disruption occurs inside the event horizon and is therefore
ignorable with respect to observables and processes outside of the
SMBH. Therefore, for the SMBH masses relevant for LISA's
bandwidth the tidal disruption of WDs is likely to be negligible
except perhaps near the plunge and merger phases for a $\lsim 10^5 M_\odot$ Schwarzschild SMBH. 


If the WD is tidally disrupted before it plunges into the SMBH, one can conceivably construct a new EFT that is valid at scales much larger than the orbital scale of the
binary in which the SMBH, the WD and the mass transferring to the
SMBH are treated collectively as an effective point particle. Given
the complexity of such a system an EFT description would be very
useful.  However, the matching procedure to determine the large
number of relevant non-minimal couplings is likely to be
difficult given that high order intrinsic multipoles will need to be
included to accurately describe this system.

We recapitulate our results from this section with an explicit statement of the Effacement Principle for {\it non-spinning} compact objects.\\

\noindent {\bf Effacement Principle for EMRIs.} {\it Tidally induced moments will affect the motion of a compact object at $O(\ve^4)$.} \\

\noindent For a white dwarf orbiting a Schwarzschild SMBH this effect may be numerically enhanced until the star undergoes tidal disruption at $O(\ve^2)$, which may be relevant when the SMBH mass is less than about $10^5 M_\odot$ for the particular example discussed here.


\section{Conclusion}

We develop an effective field theory approach for systematically
deriving the self-force on a compact object moving in an arbitrary
curved spacetime without the slow motion or weak field restrictions.
The EFT is a realization of the open quantum system paradigm  in
systems with a large scale separation such that the system's induced
fluctuations from the backreaction of the coarse-grained quantum
field is utterly negligible \cite{CalzettaHu:PRD55}. An initial value
formulation of quantum field theory is adopted here using the
closed-time-path (CTP) formalism for the in-in generating functional,
which guarantees real and causal equations of motion for the compact
object. As an illustration of the procedures involved in our approach
we showed how to derive the MST-QW equation describing the (first
order) self-force on a compact object.  The CTP formalism is needed
for a calculation of the second order self-force as will be shown in
\cite{Galley:EFT2}.

We describe the compact object as an effective point particle that is
capable of accounting for tidally induced finite size
effects. 
In calculating the effective action we encounter ultraviolet
divergences stemming from a point particle interacting with
arbitrarily high frequency modes of a graviton field. Using
Hadamard's {\it partie finie} to isolate the non-local finite part
from the quasi-local divergences we are able to implement dimensional
regularization within a (quasi-local) momentum space representation
for the graviton propagator \cite{Galley:momentum}. As such, all
power divergences can be immediately set to zero implying that only
logarithmic divergences are relevant for renormalizing the parameters
of the theory. At first order, the effective action has a power
divergence and may therefore be trivially regularized using
dimensional regularization.

In the spirit of an Effacement Principle we find that the finite size
of the compact object first affects its motion at $O(\ve^4)$ for a
non-spinning black hole and neutron star. For a white dwarf star we
deduce that such effects may be enhanced until the white dwarf is
tidally disrupted at $O(\ve^2)$ in which case the effective point
particle description, and in particular the effective field theory
developed here, breaks down. One may conceivably construct a new
effective field theory by treating the supermassive black hole, the
white dwarf and the accreting mass as an effective point
particle possessing many relevant non-minimal couplings to the
background geometry describing the intrinsic moments of this
composite object.

The leading order finite size corrections cause a deviation from the motion of a
minimally coupled point particle that is not caused by interactions
with gravitons but is due to the torques that develop on the tidally
deformed compact object. On the other hand, the self-force is
affected by the induced moments of the compact object at $O(\ve^5)$.

In summary, the EFT approach has at least two major advantages over
the existing approaches: It provides a systematic procedure for
carrying out a perturbative treatment, and an economical way to treat
the ultraviolet divergences. Our CS-EFT improves on the PN-EFT
introduced in \cite{GoldbergerRothstein:PRD73} in that it is valid
for a general curved spacetime and not limited to slow motion or weak
field conditions. These will prove to be of special benefit for
higher order self-force calculations. We will apply these steps to
calculate the self-force at second order in $\ve$ \cite{Galley:EFT2},
the gravitational radiation emitted by EMRIs \cite{Galley:EFT3} and
the self-force on spinning compact objects \cite{Galley:spin}.

\acknowledgments

We thank Eanna Flanagan, Eric Poisson, Cole Miller, Alessandra Buonanno and Ted Jacobson for useful questions and comments. CRG is gratefully indebted to Ira Rothstein for an
invitation to the Workshop on Effective Field Theory Techniques in
Gravitational Wave Physics, where much of this work had its origins.
He also thanks Ira Rothstein, Walter Goldberger and Adam Leibovich
for invaluable discussions during and after the workshop. This work
is supported in part by NSF Grant PHY-0601550.

\appendix

\section{Definitions and relations for the quantum two-point functions}
\label{app:twoptfns}

In this Appendix we collect some definitions, identities and relations for the quantum two-point functions that are relevant for this work. 

The positive and negative frequency Wightman functions are defined as
\begin{eqnarray}
    G_{\alpha \beta \gamma^\prime \delta^\prime} ^+ (x,x^\prime) &\arreq& \big\langle \hath_{\alpha \beta} (x) \hath _{\gamma^\prime \delta^\prime} (x^\prime) \big\rangle  \label{posfreq0} \\
    G_{\alpha \beta \gamma^\prime \delta^\prime} ^- (x,x^\prime) &\arreq& \big \langle \hath _{\gamma^\prime \delta ^\prime} (x^\prime) \hath _{\alpha \beta} (x) \big\rangle  ,  \label{negfreq0}
\end{eqnarray}
respectively. The angled brackets represent the quantum expectation value so that
\begin{eqnarray}
    \big\langle \hat{O} \big \rangle \equiv {\rm Tr} \Big[ \hat{\rho} (\Sigma_i) \hat{O} \Big]
\end{eqnarray}
and $\hat{\rho}(\Sigma_i)$ is the density matrix of the quantum field given on a hypersurface $\Sigma_i$ at constant coordinate time $x^0 = t _i$.

The Feynman, Dyson, Hadamard and commutator (also known as the Pauli-Jordan function or the causal function) two-point functions are, respectively,
\begin{eqnarray}
    G_{\alpha \beta \gamma^\prime \delta^\prime} ^F (x,x^\prime) &=& \big\langle T \, \hath_{\alpha \beta}(x) \hath_{\gamma^\prime \delta^\prime} (x^\prime) \big\rangle  \label{feynman0} \\
    G_{\alpha \beta \gamma^\prime \delta^\prime} ^D (x,x^\prime) &=& \big\langle T^* \, \hath_{\alpha \beta}(x) \hath_{\gamma^\prime \delta^\prime} (x^\prime) \big\rangle  \label{dyson0} \\
    G_{\alpha \beta \gamma^\prime \delta^\prime} ^H (x,x^\prime) &=& \big\langle \{ \hath_{\alpha \beta} (x), \hath _{\gamma^\prime \delta^\prime} (x^\prime) \} \big\rangle \label{hadamardfn0} \\
    G_{\alpha \beta \gamma^\prime \delta^\prime} ^C (x,x^\prime) &=& \big\langle [ \hath_{\alpha \beta} (x), \hath_{\gamma^\prime \delta^\prime} (x^\prime) ] \big\rangle  \label{commutator0}
\end{eqnarray}
where $T$ is the time-ordering operator and $T^*$ is the anti-time-ordering operator. The field commutator is independent of the particular state used to evaluate it. Given the Wightman functions in (\ref{posfreq0}) and (\ref{negfreq0}) we write the above two-point functions in the form (ignoring the tensor indices from here on)
\begin{eqnarray}
    G_F (x,x^\prime) &\arreq& \theta( t-t^\prime) G_+ (x,x^\prime) + \theta( t^\prime-t) G_- (x,x^\prime) \label{feynman1}   \nonumber \\
    && \\
    G_D (x,x^\prime ) &\arreq& \theta( t^\prime-t) G_+ (x,x^\prime) + \theta( t-t^\prime) G_- (x,x^\prime) \label{dyson1}   \nonumber \\
    &&  \\
    G_H (x,x^\prime) &\arreq& G_+ (x,x^\prime) + G_- (x,x^\prime)  \label{hadamardfn1}  \\
    G_C (x,x^\prime) &\arreq& G_+ (x,x^\prime) - G_- (x,x^\prime)  . \label{commutator1}
\end{eqnarray}
From these we define the retarded and advanced propagators by
\begin{eqnarray}
    -i G_{ret} (x,x^\prime ) &\arreq&  \theta (t-t^\prime) G_C (x,x^\prime) \label{retardedfn0} \\
    +i G_{adv} (x,x^\prime) &\arreq&  \theta(t^\prime-t) G_C(x,x^\prime)  . \label{advancedfn0}
\end{eqnarray}
In terms of the other two-point functions, these propagators satisfy the following useful identities
\begin{eqnarray}
    -i G_{ret} (x,x^\prime ) &=& G_F (x,x^\prime ) - G_- (x,x^\prime ) \\
    		&=& G_+ (x,x^\prime ) - G_D (x,x^\prime ) \\
    i G_{adv} (x,x^\prime ) &=& G_D (x,x^\prime ) - G_- (x,x^\prime ) \\
    		&=& G_+ (x,x^\prime ) - G_F (x,x^\prime )   .
\end{eqnarray}
The Feynman, Dyson and Hadamard functions are not all independent since
\begin{eqnarray}
    G_H (x,x^\prime ) &=& G_F (x,x^\prime ) + G_D (x,x^\prime ) \\
    		&=& G_+ (x,x^\prime ) + G_- (x,x^\prime )  .
\end{eqnarray}

Under the interchange of $x$ and $x^\prime$ the Feynman, Dyson and Hadamard two-point functions are symmetric, the commutator is anti-symmetric and
\begin{eqnarray}
    G_+ (x,x^\prime) &=& G_- (x^\prime, x) \\
    G_{ret} (x,x^\prime) &=& G_{adv} (x^\prime, x)
\end{eqnarray}
suggesting that the Wightman functions and retarded/advanced propagators are a kind of transpose of each other.

\section{Distributions, pseudofunctions and Hadamard's finite part}
\label{app:distro}

In this Appendix we present the basic structure, concepts and definitions of distribution theory that are relevant for this work. The reader is referred to \cite{Schwartz, Zemanian} for more information.

Consider the set of functions $\phi$ that are infinitely smooth $C^\infty$ and have compact support on any finite interval. These functions, called {\it testing} or {\it test functions}, form a set ${\cal D}$. A {\it functional} $f$ is a mapping that associates a complex number to every testing function in ${\cal D}$. A {\it distribution} is a linear and continuous functional on the space of test functions ${\cal D}$ and is frequently denoted by the symbols $\langle f, \phi \rangle$ and $f$.

For a locally integrable function $f(t)$ we can associate a natural distribution through the convergent integral
\begin{eqnarray}
    \big\langle f, \phi \big\rangle \equiv \int_{-\infty}^\infty dt \, f(t) \phi(t)
\end{eqnarray}
for some testing function $\phi \in {\cal D}$. Notice that we are using the same symbol to denote both the distribution and the function that generates the distribution. This is an example of a regular distribution. All distributions that are not regular are {\it singular distributions} and will be our main concern in the rest of this Appendix. An example of a singular distribution is the well known delta functional $\delta$. As this is not generated by a locally integrable function $\delta(t)$ (even as the limit of a sequence of locally integrable functions \cite{Zemanian}) it must be a singular distribution.

Often, a singular distribution gives rise to a singular integral, which can be written in terms of its divergent and finite parts. For the purposes of clarity and illustration it is best to consider a simple example. Let us compute the integral
\begin{eqnarray}
    \Big\langle \frac{\theta(t)}{ t } , \phi \Big\rangle = \int_0 ^\infty dt \, \frac{\phi(t) }{ t }   \label{distroex0}
\end{eqnarray}
for $\phi(t)$ a testing function in ${\cal D}$ and $\theta(t)$ the step, or Heaviside, function. This integral is obviously divergent since $1/t$ is not a locally integrable function at the origin. Nevertheless, we may extract the finite part (in the sense of Hadamard \cite{Hadamard}) of the integral by isolating the divergences from the finite terms.

To this end we write
\begin{eqnarray}
    \phi(t) = \phi(0) + t \, \psi(t)   \label{phipsidef0}
\end{eqnarray}
where $\psi(t)$ is a continuous function for all $t$. Putting this into (\ref{distroex0}) and integrating gives
\begin{eqnarray}
    \Big\langle \frac{ \theta(t) }{ t} , \phi \Big\rangle &=& \lim _{\epsilon \to 0^+} \Bigg[ \phi(0) \log b - \phi(0) \log \epsilon + \int _\epsilon ^b dt \, \psi(t) \Bigg]   \nonumber \\
    &&
\end{eqnarray}
where we assume that the testing function $\phi(t)$ vanishes for $t \ge b$ for some real number $b$. The finite part of (\ref{distroex0}) is defined to be the remainder upon subtracting off the divergent contribution(s). In this case, dropping the $\log \epsilon$ term gives the finite part of the integral
\begin{eqnarray}
    Fp \int_0 ^\infty dt \, \frac{ \phi(t) }{ t} = \phi(0) \log b + \int _0 ^\infty dt \, \psi(t)  \label{Fp0}
\end{eqnarray}
where the symbol $Fp$ denotes the Hadamard finite part of the integral. Therefore, the divergent part of the integral is given by $ - \phi(0) \log \epsilon$.

A distribution that generates the finite part of the integral is called a {\it pseudofunction}, which we now calculate for this example. Inserting (\ref{phipsidef0}) into the finite part (\ref{Fp0}) gives
\begin{eqnarray}
    Fp \int_0 ^\infty dt \, \frac{ \phi(t) }{ t } = \lim _{\epsilon \to 0^+} \Bigg[ \int _\epsilon ^\infty dt \, \frac{ \phi(t) }{ t } +  \phi(0) \log \epsilon \Bigg]   . ~~
\end{eqnarray}
Since
\begin{eqnarray}
    \phi(0) = \int_{-\infty}^\infty dt \, \delta(t) \phi(t) = \langle \delta, \phi \rangle
\end{eqnarray}
it follows that the finite part can be written as an integral of a distribution with a testing function
\begin{eqnarray}
    Fp \int_0 ^\infty dt \, \frac{ \phi(t) }{ t } = \lim _{\epsilon \to 0^+}  \int _\epsilon ^\infty dt \Bigg[ \frac{ 1 }{ t } + \delta(t) \log \epsilon \Bigg] \phi(t)  , ~~
\end{eqnarray}
which defines the pseudo-function,
\begin{eqnarray}
    Pf \frac{ \theta(t) }{t} = \frac{\theta (t) }{t}  + \delta (t) \lim_{\epsilon \to 0^+} \log \epsilon   .
\end{eqnarray}
Therefore, the finite part of the integral generates a pseudo-function (a regular distribution) that yields a finite value when integrated with a testing function
\begin{eqnarray}
    \int _{-\infty} ^\infty dt \, Pf \frac{\theta(t) }{t} \, \phi(t) = Fp \int _{-\infty}^\infty dt \, \frac{\theta(t)}{ t} \, \phi(t) .
\end{eqnarray}

Quite generally, the value that the distribution assigns to a testing function will have a divergent part consisting of both power divergences and powers of logarthmically diverging terms so that
\begin{eqnarray}
    I (\epsilon) = \sum_{p=1} ^N \frac{ a_p }{ \epsilon^p } + \sum_{p=1} ^M b_p \log ^p \epsilon
\end{eqnarray}
for some appropriate integers $N, M$. This form for $I(\epsilon)$ is related to the so-called Hadamard's ansatz \cite{Hadamard} and appears often in regularizing divergent quantities involving two-point functions of a quantum field in curved spacetime \cite{BirrellDavies, Wald}.

\section{Dimensional regularization of effective action}
\label{app:regularize}

In this Appendix we give the explicit calculation of the divergent part of the $O(\ve)$ effective action. In Riemann normal coordinates, $y^\hata$ describes the coordinate of a point $x^\prime$ relative to the origin at $x$ and is defined in terms of a tetrad $e_\alpha^\hata$ at $x$ through
\begin{eqnarray}
	 y^\hata = - e^\hata _{\alpha} (x) \, \sigma^{\alpha} ( x, x^\prime)  . \label{defRNC0}
\end{eqnarray}
Here $\sigma(x,x^\prime)$ is Synge's world function, which numerically equals half the squared geodesic interval between $x$ and $x^\prime$, and $\sigma^\alpha \equiv \sigma^{;\alpha}$ is proportional to the vector at $x$ that is tangent to the unique geodesic connecting $x$ and $x^\prime$. See \cite{Poisson:LRR} for further details. Let us equate the points $x$ and $x^\prime$ with $z_+^\mu (\tau)$ and $z_+^\mu (\tau^\prime)$, respectively, on the leading order (i.e., geodesic) particle worldline. 

The arbitrariness of the tetrad at $x$ implies that we may choose
\begin{eqnarray}
	e^{\hat{0}}_\alpha (\tau) = - u_{+ \alpha} (\tau)  .
\end{eqnarray}
if we maintain the condition $e^\hata _\alpha e^\alpha_{\hatb} = \delta^a _b$.
Hence, the particle's 4-velocity is orthogonal to the components of the tetrad in the spatial directions, $e^\hati_\alpha u^\alpha_+ = 0$ where $\hati = 1, \ldots, d-1$. Using these relations and (\ref{defRNC0}), the Riemann normal coordinate representation of the point $x^\prime = z_+^\mu (\tau^\prime)$ on the geodesic worldline is given by
\begin{eqnarray}
          y^\hata  = e^\hata _{\alpha} (\tau) \, u_+^\alpha (\tau) ( \tau- \tau^\prime) = - \delta^\hata_{\hat{0}} \, (\tau - \tau^\prime)  
\end{eqnarray}
implying that
\begin{eqnarray}
    k \cdot y = k_\hata y^\hata =  - k_\hatzero \, (\tau - \tau^\prime)  .
       \label{kdoty}
\end{eqnarray}

Integrating the divergent integral in (\ref{divintegral01}) over $s=\tau^\prime - \tau$ gives a delta function forcing $k_\hatzero$ to vanish so that
\begin{eqnarray}
   I ^{\hatm \hatn} (\tau) &=& - \frac{i}{2} \, \frac{d-3}{d-2} w^{\hatm \hatn} \int _{\bk}  \frac{ 1 }{ \bk^2 } \nonumber \\
    && + \frac{8i}{3} \, \frac{d^2 - 6d +10}{ (d-2)^2 } \int_{\bk} \frac{ k^\hati k^\hatj }{ \bk^6 } \, R_{\hata \hati \hatzero \hatj} \eta^{\hatm ( \hatn} \delta^{\hata ) }_\hatzero \nonumber \\
   && \label{divintegral3}
\end{eqnarray}
where $\bk$ is the $d-1$ dimensional spatial momentum with $\int_{\bk} = \int d^{d-1}k / (2\pi)^{d-1}$ and $\hati, \hatj = 1, \ldots , d-1$ are indices for the spatial directions. 

The momentum integral in the first line of (\ref{divintegral3}) can be integrated by giving a small mass $m_g$ to the graviton so that
\begin{eqnarray}
    && {\hskip-0.5in} \int _{\bk}  \frac{ 1 }{ ( \bk ^2 + m_g^2  )^\alpha }   \nonumber \\
    &=& \frac{ 2 \pi^{(d-1) / 2} }{ (2\pi)^{d-1} \Gamma \left( \frac{ d-1}{2} \right) } \int _0 ^\infty dk \, \frac{ k^{d-2} }{ (k^2+m_g^2 ) ^\alpha } \label{divintegralk0} \\
     &=& \frac{1}{ (4 \pi)^{(d-1)/2} } \, \frac{ \Gamma \left( \alpha + \frac{ 1-d }{ 2} \right) }{ \Gamma (\alpha) } \, (m_g^2) ^{ \frac{d-1}{2} -\alpha }  ,  \label{divintegralk1}
\end{eqnarray}
for some positive integer $\alpha$. Strictly speaking, the integral in (\ref{divintegralk0}) does not necessarily converge for $d=4$. However, by analytically continuing to other values for $d$ we find that the integral can be made to converge. In this way, the divergence is renormalized via the analytic continuation and a finite result is obtained upon letting $m_g \to 0$ and then choosing $d=4$. For $\alpha=1$ and $d=4-\ve$ the integral is
\begin{eqnarray}
    \int_{\bk} \frac{ 1 }{ \bk ^2 + m_g^2  } = - \frac{ m_g }{4 \pi} + O(\ve)   . \label{divintegralk2}
\end{eqnarray}
In the second line of (\ref{divintegral3}) we find that
\begin{eqnarray}
	R_{\hata \hati \hatzero \hatj} \int_{\bk} \frac{ k^\hati k^\hatj}{ \bk^6 } = R_{\hata \hati \hatzero} ^{~~~\,\hati} \int_{\bk} \frac{1}{\bk ^4} = R_{\hata \hatzero} \int_{\bk} \frac{1}{\bk^4} , \label{divintegral4}
\end{eqnarray}
which vanishes because the background spacetime is vacuous. 

Therefore, in the limit that the graviton mass goes to zero we find that (\ref{divintegralk2}) vanishes and
\begin{eqnarray}
    I ^{\mu \nu} (\tau) = e^\mu_\hatm e^\nu_\hatn  I ^{\hatm \hatn} (\tau) = 0  ,
\end{eqnarray}
as claimed.


\bibliographystyle{physrev}
\bibliography{eft1_v2}

\setlength{\parskip}{1em}



\end{document}